\documentclass[a4paper,fleqn,usenatbib]{mnras}

\usepackage{newtxtext,newtxmath}
\usepackage[T1]{fontenc}
\usepackage{ae,aecompl}
\usepackage{booktabs}

\usepackage{graphicx}	% Including figure files
\usepackage{amssymb}% Extra maths symbols
\usepackage{amsmath}	% Advanced maths commands
\usepackage[dvipsnames]{xcolor}    % can be removed once comments are removed
\usepackage[super]{nth}
\usepackage{microtype}

\newcommand*{\qv}{$q_{\rm V}(z)$}

\newcommand*{\LCDM}{$\Lambda$CDM}
\newcommand*{\La}{$\Lambda$}
\newcommand*{\kmsMpc}{kms$^{-1}$Mpc$^{-1}$}

\title[Constraints on  the  interacting vacuum scenario]{Constraints on  the  interacting vacuum -- geodesic CDM scenario}
\author[M. Martinelli et al.]{
Matteo Martinelli,$^{1}$\thanks{E-mail: martinelli@lorentz.leidenuniv.nl}
Natalie B. Hogg,$^{2}$
Simone Peirone,$^{1}$
Marco Bruni,$^{2,3}$
David Wands$^{2}$
\\
$^{1}$Institute Lorentz, Leiden University, PO Box 9506, Leiden 2300 RA, The Netherlands \\
$^{2}$Institute  of  Cosmology  and  Gravitation,  University  of  Portsmouth, Burnaby  Road,  Portsmouth  PO1  3FX, UK\\
$^3$ INFN Sezione di Trieste, Via Valerio 2, 34127 Trieste, Italy
}

\date{Accepted 2019 July 4. Received 2019 June 21; in original form 2019 March 1}

\pubyear{2019}

\begin{document}
\label{firstpage}
\pagerange{\pageref{firstpage}--\pageref{lastpage}}
\maketitle

\begin{abstract}
We investigate an interacting dark sector scenario in which the vacuum energy is free to interact with cold dark matter (CDM), which itself is assumed to  cluster under the sole action of gravity, i.e.\ it is in free fall (geodesic), as in $\Lambda$CDM. The interaction is characterised by a dimensionless coupling \qv, in general a function of redshift. Aiming to reconstruct the evolution of the coupling, we use CMB data from Planck 2015, along with baryon acoustic oscillation, redshift space distortion and Type Ia supernova measurements  to constrain various parametrizations of $q_{\rm V}(z)$. We present the full linear perturbation theory of this interacting scenario and use MCMC sampling to study five different cases: two cases in which we have \LCDM~evolution in the distant past, until a set redshift $z_{\rm trans}$, below which the interaction switches on and $q_{\rm V}$ is the single sampled parameter, with $z_{\rm trans}$ fixed at $z_{\rm trans}=3000$ and $z_{\rm trans}=0.9$ respectively; a case where we allow this transition redshift to vary along with $q_{\rm V}$; a case in which the vacuum energy is zero for  $z>z_{\rm trans}$ and then begins to grow once the interaction switches on; and the final case in which we bin \qv~in four redshift bins to investigate the possibility of a dynamical interaction, reconstructing the redshift evolution of the function using Gaussian processes. We find that, in all cases where the high redshift evolution is not modified, the results are compatible with a vanishing coupling, thus finding no significant deviation from \LCDM.

\end{abstract}


%%%%%%%%%%%%%%%%%%%%%%%%%%%%%%%%%%%%%%%%%%%%%%%%%%

%%%%%%%%%%%%%%%%% BODY OF PAPER %%%%%%%%%%%%%%%%%%

\section{Introduction} \label{sec:intro}
Over the past 20 years, observational cosmology has provided a wealth of evidence in support of the idea that the expansion of the Universe is accelerating. The first direct evidence for the acceleration came from  Type Ia supernovae observations \citep{Riess1998,Perlmutter1999} and subsequent measurements of the cosmic microwave background (CMB) \citep{Hinshaw2009, Aghanim:2018eyx} and other cosmological probes such as baryon acoustic oscillations (BAOs) \citep{Alam:2016hwk} have all confirmed the late-time dominance of a dark energy component in our Universe.

The standard cosmological model, \LCDM, has been largely successful in explaining these measurements, with the cosmological constant, \La, being the simplest driver of an accelerated expansion and cold dark matter (CDM) being responsible for structure formation. However, there are problems with  \LCDM \ which motivate the investigation of alternative models. These problems manifest in both the discrepancy between the predicted and observed values of the cosmological constant \citep{Weinberg1989, Adler1995}, and in the tensions that exist between low redshift probes of the expansion rate and structure growth and the corresponding values inferred from CMB measurements (for which a cosmological model must be assumed) \citep{Macaulay:2013swa, Bernal:2016gxb}.

In recent years, the precision of surveys has improved and these tensions have become more apparent, particularly in the value of the Hubble parameter today, $H_0$; the most recent CMB measurement, from the Planck satellite, is $H_0 = 67.4 \pm 0.5$ \kmsMpc \ \citep{Aghanim:2018eyx}, whereas the most recent local determination, from the Hubble Space Telescope, is $H_0 = 73.45 \pm 1.66$ \kmsMpc \ \citep{Riess2018}, a discrepancy of $3.7 \sigma$. Other distance ladder-independent probes do not seem to ease the tension, with the LIGO measurement of $70^{+12.0}_{-8.0}$ \kmsMpc \ \citep{Abbott:2017xzu} and a recent H0LiCOW quadruple lensed quasar measurement of $72.5^{+2.1}_{-2.3}$ \kmsMpc \ \citep{Birrer2018} falling between the CMB and distance ladder results. 

The tension in $\sigma_8$, the amplitude of the linear matter power spectrum on a scale of $8 h^{-1}$Mpc, is less severe than that in $H_0$, but is yet another indicator of problems with \LCDM. Once again, the discrepancy appears between measurements of $\sigma_8$ at large and small scales, most noticeably the scales probed by the CMB and the smaller scale indicators of large scale structure (LSS), such as galaxy cluster counts, weak lensing and redshift space distortion (RSD) measurements \citep{Battye:2014qga}, with LSS giving a lower value than CMB \citep{Abbott:2017wau}.

It remains to be seen whether these tensions will survive as the new generation of surveys, satellites and telescopes begins to provide us with data and new analysis techniques are developed. An interesting example of such a novel method has latterly been described in the literature, with compelling results \citep{Aubourg2015, Macaulay:2018fxi}. These authors invert the distance ladder, anchoring the Type Ia supernova measurements to BAOs rather than to the parallax distances of Cepheid variable stars. Using this method, along with 207 new DES supernovae, \cite{Macaulay:2018fxi} find a value of $H_0 = 67.77 \pm 1.30$ \kmsMpc, which is in excellent agreement with the derived value from Planck.  This hints at an uncertainty in the measuring of parallax distances which could be leading to a miscalibration of the distances to the Cepheids. This uncertainty could be reduced by future data from Gaia \citep{Beaton:2018fyo} and LSST \citep{Ivesic2012}. However, observational advances are just one way the $H_0$ and $\sigma_8$ tensions could be resolved; an alternative is to examine new theoretical models of dark energy.

In this work, we explore the phenomenology of a scenario in which the vacuum energy is free to interact with dark matter. The idea of a decaying vacuum energy as been afforded a great deal of study in the literature (see, for example, \cite{Bertolami1986, Pavon1991, Alrawaf1996, Shapiro:2000dz, Sola:2011qr, Wands:2012vg}) and other dynamical and interacting dark energy models have also been investigated, often with the conclusion that not only can cosmological tensions be relieved in such models, but they may even be favoured over \LCDM \ (see, for example, \cite{Salvatelli2014, Wang:2015wga, Zhao:2017cud, Sola:2017lxc, DiValentino:2017iww, Kumar:2017dnp, Sola2018:lcd, Yang:2018qmz, Wang:2018fng} for more details). 

The specific scenario we here consider retains general relativity as the description of gravity, while allowing for a possible exchange of energy between cold dark matter and the vacuum, i.e.\ a dark energy with an equation of state parameter $w=-1$ \citep{1931Natur.127..706L,Lemaitre1933}. This scenario does not introduce any additional dynamical degrees of freedom with respect to \LCDM~ \citep{Wands:2012vg}. The interaction allows for the energy density of the vacuum, $V$, to change, while CDM can freely cluster under the sole action of gravity i.e.\ CDM remains geodesic, as in $\Lambda$CDM. We investigate the possibility of such an interaction by choosing a simple parametrization and studying its behaviour as a function of redshift. As we show in Section \ref{sec:theory}, the interaction is described -- in the synchronous comoving gauge and under the assumption of geodesic CDM -- by a single background function $Q(z)$ which we model as $Q= q_{\rm V} H V$, where \qv\ is a dimensionless function. Based on this, we consider five different cosmologies, with the general aim of reconstructing \qv\  with step functions in different redshift bins, using the values $q_{\rm V}$ has in each bin as parameters.  In particular, a four bins case is  essentially model-independent.

The first  two cosmologies, which we name \textit{Cfix}, consider a physical scenario in which we have a \LCDM \ evolution in the past up to a fixed transition redshift $z_{\rm trans}$. At $z$  lower than $z_{\rm trans}$, the interaction switches on and the vacuum energy starts to evolve. The two cases differ in the redshift of the transition: in the first we assume that the interaction starts at high redshift, with $z_{\rm trans}=3000$; in the other we assume  $z_{\rm trans}=0.9$, in order to compare with the same case considered by \cite{Salvatelli2014}. For these two \textit{Cfix} cases we sample over the usual cosmological parameters, with the addition of the single interaction parameter, $q_{\rm V}$. 

The third case, \textit{Cvar}, is similar to the first two, but we additionally sample over the transition redshift, $z_{\rm trans}$. The fourth case, which we call \textit{seeded vacuum energy} or \textit{SVE}, mimics a physical scenario in which the coupling causes the vacuum energy to suddenly grow from zero up to a  `seed' value at $z_{\rm trans}$. At lower redshifts, the interaction then behaves as in the previous three cases i.e.\ with a constant $q_{\rm V}$, and the vacuum evolves accordingly. Therefore this case, like the third, has two free parameters: $q_{\rm V}$ and $z_{\rm trans}$. The fifth case we consider is the model-independent scenario in which we allow the interaction to evolve in four redshift bins, using four different values of the interaction strength $q_{\rm V}$. We call this the \textit{4bins} case. We use MCMC techniques to constrain the coupling and transition redshifts in each case, using the latest observational datasets.

We draw attention to the previous work of some of the authors, \cite{Salvatelli2014}, and wish to emphasize the differences between that Letter and the current paper. In this work, we make use of the new data that is now available, especially the Planck 2015 likelihood, along with new baryon acoustic oscillation, redshift space distortion and Type Ia supernova data. We also use a less restrictive prior on the coupling parameter in our parameter inference, allowing us to investigate the possibility of an energy transfer both from dark matter to the vacuum and vice versa. We will discuss this further in Section \ref{sec:rebuttal}.

The rest of the paper is organized as follows: in Section \ref{sec:theory} we present the theory of the interacting vacuum scenario, including the equations governing the evolution of the background and perturbations. In Section \ref{sec:qreconst} we outline the parameterization of the interaction and the methods of reconstruction studied in this work. In Section \ref{sec:analysis} we discuss the data and analysis methods used in our investigation and in Section \ref{sec:results} we present our results, followed by a discussion in Section \ref{sec:discussion}. We make some comments on other recent works in this area in Section \ref{sec:rebuttal} and we finally conclude with Section \ref{sec:conclusions}. 

\section{Cold dark matter -- vacuum energy interaction} \label{sec:theory}
In this section we outline the theoretical framework for the interacting vacuum scenario, beginning with a summary of the general covariant theory and progressing to the details of the scenario in a Friedman-Lema\^{i}tre-Robertson-Walker background with perturbations. See \cite{Wands:2012vg} for more details.

\subsection{Covariant theory of the interacting vacuum} 
In \LCDM, the cosmological constant \La \ represents the vacuum energy of the Universe, and in a classical sense, this vacuum energy can be treated as a non-interacting perfect fluid with an equation of state parameter $w=-1$, as was realised by Lema\^{i}tre \citep{1931Natur.127..706L,Lemaitre1933}. 

The energy--momentum tensor of a perfect fluid is
\begin{equation}
T^\mu_\nu = P g^\mu_\nu + (\rho + P) u^\mu u_\nu, \label{eq:emperf}
\end{equation}
where $\rho$ is the energy density, $P$ the pressure and $u^\mu$ the 4-velocity of the fluid. 

We define the energy--momentum tensor of the vacuum as
\begin{align}
\check{T}^\mu_\nu = -V g^\mu_\nu,
\end{align}
and by comparison with \eqref{eq:emperf} we can identify $V = -\check{P} = \check{\rho}$, i.e. $V$ is the vacuum energy density. This means that the equation of state parameter $w = P/\rho$  is equal to $-1$, as it is for the cosmological constant \La. Moreover, this form of the vacuum energy--momentum tensor leaves the vacuum 4-velocity undefined and any 4-vector is an eigenvector of $\check{T}^\mu_\nu$. Therefore all observers measure the same vacuum energy density $V$; in other words, the vacuum energy is boost invariant. In the following, $u^\mu$ therefore denotes the 4-velocity of cold dark matter (CDM).

Denoting the energy--momentum tensor of CDM with $T^\mu_\nu$ and its energy density with $\rho_c$,
\begin{align}
 T^\mu_\nu = \rho_c u^\mu u_\nu,
\end{align}
we can introduce an interaction between CDM and the vacuum energy in the following way:
\begin{align}
\nabla_\mu T^\mu_\nu &= -Q_\nu \label{eq:gencdmcons}, \\
\nabla_\mu \check{T}^\mu_\nu &= -\nabla_\nu V = Q_\nu, \label{eq:genvaccons}
\end{align}
where the interaction 4-vector $Q^\nu$ represents the energy--momentum flow between vacuum and CDM. 

If $T^{\mu\nu}_{\rm tot}= T^{\mu\nu} + \check{T}^{\mu\nu}$ is the total energy--momentum tensor, then the form of the interaction in \eqref{eq:gencdmcons} and \eqref{eq:genvaccons} ensures the total conservation equation $\nabla_\mu T^{\mu\nu}_{\rm tot} =0$, which, in Einsteinian gravity follows from the Bianchi identity $\nabla_\mu G^{\mu\nu} \equiv 0$. We note that this scenario reduces to the standard $\Lambda$CDM case when $Q^\nu=0$, as this implies $V=$ constant.

We can project the interaction 4-vector in two parts parallel and orthogonal to the CDM 4-velocity:
\begin{align}
Q^\mu &= Q u^\mu + f^\mu, \label{eq:coupling}
\end{align}
where, in the frame of observers comoving with CDM, $Q$ represents the energy flow, and  $f^\mu$  the momentum exchange between CDM and  vacuum; $f^\mu$ is orthogonal to $u^\mu$ i.e.\ $f^\mu u_\mu =0$.

Following \cite{Salvatelli2014} and \cite{Wang:2014xca}, we consider the simplest case of interaction: a pure energy exchange in the CDM frame wherein $f^\mu = 0$, and so $Q^\mu = Qu^\mu$. The 4-force, $f^\mu$, is related to the 4-acceleration $a^\mu = u^\alpha \nabla_\alpha u^\mu$ by
\begin{align}
f^\mu &= a^\mu \rho_c.
\end{align}
Since we set $f^\mu =0$, it follows that $a^\mu = 0$, meaning there is no acceleration of CDM due to the interaction and hence CDM remains geodesic. We may call this interacting scenario the \textit{geodesic CDM scenario} (see also \cite{PhysRevD.87.083503}). It follows from this geodesic CDM assumption that the effective sound speed of matter perturbations is zero and hence the Jeans length is  also zero, meaning that there is no damping of matter perturbations on scales smaller than the Jeans length. However, the interaction will still affect structure growth, as discussed below in subsection \ref{subsec:pert}. 

A second important consequence of the assumption of pure energy exchange is that, following \eqref{eq:genvaccons}, the CDM 4-velocity $u^\mu$ consequently defines a potential flow and the CDM fluid is thus irrotational \citep{Borges:2017jvi}. This is a sufficient approximation of the behaviour of cold dark matter at early times and on large scales, in a linear regime where only scalar perturbations are relevant for structure formation,   but at late times it is expected that non-linear structure growth will lead to vorticity. At late times, dark matter haloes are rotationally supported and in this non-linear regime, a gravito-magnetic frame-dragging vector field is generated \citep{Bruni:2013mua}. Dark matter composed of a purely irrotational fluid would have strong observational signatures (in particular, the rapid formation and growth of supermassive black holes \citep{Sawicki2013}), so our assumption of the pure energy exchange which allows CDM to remain geodesic must break down below some length scale. Further investigation of this limit is left to a future work.

\subsection{Flat FLRW background}
In a spatially flat Friedmann--Lema\^{i}tre--Robertson--Walker background, \eqref{eq:gencdmcons} and \eqref{eq:genvaccons} reduce to the coupled energy conservation equations,
\begin{align}
\dot{\rho}_c + 3H\rho_c &= -Q, \label{eq:flrwcdm} \\
\dot{V} &= Q, \label{eq:flrwvac}
\end{align}
where $H$ is the Hubble expansion function and $Q$ is the interaction term. 

\subsection{Linear perturbations} \label{subsec:pert}
We now consider the linear, scalar perturbations about the FLRW metric. With the inclusion of these, the line element in a general gauge becomes
\begin{align}
ds^2 = &-(1+2\phi)dt^2 + 2 a\partial_i B dx^i dt \nonumber \\ &+ a^2 [(1-2\psi) \delta_{ij} + 2\partial_i \partial_j E] dx^i dx^j. 
\end{align}
The perturbed energy density of cold dark matter is given by $\rho_c + \delta \rho_c$, and the perturbed 4-velocity of matter is
\begin{align}
&u^\mu = [1 - \phi, \ a^{-1} \partial^i v],\\
&u_\mu = [-1-\phi, \ \partial_i \theta], \label{eq:norm}
\intertext{where}
&\partial^i v = a \frac{\partial x^i}{\partial t}, \\
& \theta = a(v + B).\label{eq:theta}
\end{align}

In the geodesic CDM scenario, where in \eqref{eq:coupling} $f^\mu = 0$, the perturbed energy conservation equations for CDM and the vacuum become
\begin{align}
&\delta \dot{\rho}_c + 3H \delta \rho_c - 3\rho_c \dot{\psi} + \rho_c \frac{\nabla^2}{a^2} (\theta + a^2 \dot{E} - aB) = - \delta Q - Q \phi, \label{eq:pertenergy1}\\
&\delta\dot{V} = \delta Q + Q \phi \label{eq:pertenergy2},
\intertext{and the momentum conservation equations become}
&\dot{\theta} + \phi = 0, \label{eq:pertmom1}\\
&-\delta V = Q \theta \label{eq:pertmom2}. 
\end{align}
Considering that we are interested in the geodesic CDM scenario, with the  interaction consisting of a pure energy exchange in the CDM frame, i.e.\  $Q^\mu=Qu^\mu$, the CDM 4-velocity $u^\mu$  acquires a central role, and it is therefore useful to consider a velocity-orthogonal slicing where $u^\mu$ coincides with  the normal to the constant-time hypersurfaces \citep{1984PThPS..78....1K,Malik:2008tq}. 

In  this case the spatial components of $u_\mu$ in \eqref{eq:norm} vanish, and so $\theta =0$, which then implies  $v+B=0$  from \eqref{eq:theta}. The main convenience of this time-slicing with $\theta =0$ is that the vacuum is spatially homogeneous on these hypersurfaces,  $\delta V =0$, which follows from \eqref{eq:pertmom2}. In this slicing, we can then specify a gauge.

A convenient choice of gauge for the numerical analysis discussed later is the synchronous gauge comoving with the 4-velocity of CDM, fixed by setting $\phi = v = B = 0$. With this choice, \eqref{eq:pertmom1} becomes an identity, \eqref{eq:pertmom2} again implies $\delta V =0$ and \eqref{eq:pertenergy2} therefore gives $\delta Q =0$: both the interaction and the vacuum are spatially homogeneous with this gauge choice. The interaction therefore does not explicitly appear in the perturbation equations \eqref{eq:pertenergy1}, \eqref{eq:pertenergy2} and it is not necessary to evolve the vacuum perturbations once this choice of gauge is made\footnote{We wish to emphasize that the vacuum \textit{is} perturbed in a general spacetime sense; it is only homogeneous in the frame of observers comoving with  the geodesic CDM, where $\delta V=0$.}.

However, it is usual to use the density contrast $\delta_c = \delta \rho_c /\rho_c$ to describe matter perturbations. In doing so, the interaction is reintroduced via the evolution of $\rho_c$ from \eqref{eq:flrwcdm}. We find that $\delta_c$ evolves as
\begin{align}
\dot{\delta}_c = \frac{Q}{\rho_c}\delta_c + 3 \dot{\psi} - \nabla^2 \dot{E} \label{eq:densitycontrast}.
\end{align}
This point cannot be stressed enough, as it shows that the interaction has an effect on the perturbations and not just the background. This has important implications for cosmological structure growth, as we will further describe in subsection \ref{subsec:rsd}.

One may feel that the discussion of perturbations in CDM and the vacuum only is too idealised, especially considering that in our numerical analysis described in Section \ref{sec:analysis}, we make use of the Einstein-Boltzmann code \texttt{CAMB}~\citep{Lewis:1999bs,Howlett:2012mh} in which baryons and radiation are also included. In such a multi-component case a common gauge choice is that of the total-matter gauge, with a 4-velocity chosen to be the eigenvector of the total energy-momentum tensor \citep{1984PThPS..78....1K}. In such a gauge the CDM would have a peculiar velocity and both the vacuum and the interaction would be inhomogeneous. However, \texttt{CAMB} works in the synchronous gauge comoving with CDM and therefore the perturbation equations of the other components remain unchanged when one modifies \texttt{CAMB} from its basic $\Lambda$CDM version. This greatly simplifies the analysis of the geodesic CDM scenario we consider in this paper.

\subsection{Redshift space distortions in interacting cosmologies} \label{subsec:rsd}
An interacting scenario such as the one described above has a non-trivial effect on the growth of structure, as we will now explain. The peculiar velocities of galaxies, $\mathbf{v}$, cause a stretching and squashing in their shapes when plotted in redshift space. In $\Lambda$CDM, where there is no interaction, these redshift space distortions constrain structure growth because the divergence of the peculiar velocity field, $\mathbf{\nabla} \cdot \mathbf{v}$, is related to the time derivative of the density contrast,
\begin{align}
\dot{\delta}_c = -\frac{1}{a} \mathbf{\nabla} \cdot \mathbf{v} \label{eq:divv}\;.
\intertext{One can write this time derivative  in terms of a growth factor $f$ as}
\dot{\delta}_c = - \delta_c H f,
\intertext{where  $f$ is defined as}
f \equiv \frac{d \ln D}{d \ln a},
\end{align}
and where $D$ is the amplitude of the linear growing mode \citep{Hamilton2001}. These distortions therefore allow a constraint to be placed on the growth rate of structure in the form of $f \sigma_8$, where $\sigma_8$ is the amplitude of the linear matter power spectrum on a scale of $8h^{-1}$Mpc. Equation \eqref{eq:divv} can be interpreted in relativistic perturbation theory as relating $\delta_c$ in the comoving-synchronous gauge of the previous section to $\mathbf{\nabla} \cdot \mathbf{v}$ in the Newtonian--Poisson gauge \citep{1984PThPS..78....1K, Malik:2008tq}.

However, in the interacting vacuum scenario, the interaction enters into the equation for the evolution of the density contrast, \eqref{eq:densitycontrast}. Relating the $\mathbf{\nabla} \cdot \mathbf{v}$ term with the metric perturbations in the synchronous comoving gauge gives
\begin{align}
\mathbf{\nabla} \cdot \mathbf{v} \equiv -a(3 \dot{\psi} - \nabla^2 \dot{E}),
\intertext{and so}
\mathbf{\nabla} \cdot \mathbf{v} = - a \dot{\delta}_c + \frac{a Q \delta_c}{\rho_c},\\
\mathbf{\nabla} \cdot \mathbf{v} = -a \delta_c H f_i,
\intertext{where $f_i$ is the modified growth rate in the interacting vacuum cosmology,}
f_i = f - \frac{Q}{H \rho_c}.
\label{eq:fi}
\end{align}
This means that in the interacting vacuum scenario, the redshift space distortions that we observe place a constraint on a new parameter that we may call $f_i \sigma_8$. This has been studied in \cite{Borges:2017jvi}, and a similar effect in a cosmology with a scalar field that conformally and disformally couples to dark matter was noted in \cite{Kimura:2017fnq}. 

An unmodified version of the code \texttt{CAMB} would compute the parameter $f \sigma_8$ as written in Equation 33 of \cite{Ade:2015xua},
\begin{align}
f \sigma_8 (z) \equiv \frac{\left[\sigma_8^{(\rm vd)}(z)\right]^2}{\sigma_8^{(\rm dd)}(z)} \label{eq:planckfs8},
\end{align}
where $\sigma_8^{(\rm vd)}$ is the smoothed density-velocity correlation and $\sigma_8^{(\rm dd)}$ the smoothed density autocorrelation. The peculiar velocity in \eqref{eq:planckfs8} is the Newtonian--Poisson gauge velocity of the baryons and CDM. However, as we will explain in Section \ref{sec:analysis}, we modify \texttt{CAMB} to include our interacting scenario. It follows that the modified \texttt{CAMB} actually computes the right hand side of \eqref{eq:planckfs8}, which we may interpret as the parameter $f_i \sigma_8$. We can therefore safely use redshift space distortion data when attempting to constrain the interaction strength. However, this is not a direct constraint on the growth factor, $f$.

\section{Coupling function reconstruction}\label{sec:qreconst}
In order to constrain the interaction with available data, we write the covariant coupling in \eqref{eq:coupling} as
\begin{equation}\label{eq:inter}
Q = -q_{\rm V} \frac{1}{3} \Theta V,
\end{equation}
where $\Theta = \nabla_\mu u^\nu$ is the expansion scalar and $q_{\rm V}$ is a dimensionless function that represents the strength of the coupling.

In an FLRW background, \eqref{eq:inter} reduces to
\begin{equation}\label{eq:bigQ}
Q(z) = -q_{\rm V}(z) H(z) V(z),
\end{equation}
and hence the energy conservation equations \eqref{eq:flrwcdm} and \eqref{eq:flrwvac} become
\begin{align}
\dot{\rho}_c + 3H\rho_c &= q_{\rm V}(z) H(z) V(z), \label{eq:qcdm} \\
\dot{V} &= -q_{\rm V}(z) H(z) V(z) \label{eq:qvac}.
\end{align}

Now that we have the differential equations written in terms of the dimensionless coupling \qv, we need to model the evolution of this function in redshift in terms of some numerical parameter that we will later constrain with cosmological data. 

In this paper, however, we are aiming to reconstruct the coupling rather than test specific models, adopting an agnostic standpoint regarding \qv\ and letting the data to tell us what this function is likely to be.  The simplest way to  proceed is to use step functions, approximating the coupling function \qv\ with one or more constant values of $q_{\rm V}$ in a series of redshift bins.

\footnote{Notice that adopting a step function reconstruction for \qv\ introduces  discontinuities in $\dot{\rho}_c$ and $\dot{V}$ in \eqref{eq:qcdm} and \eqref{eq:qvac} at the boundaries of the redshift bins; however this is not a problem, as the resulting $\rho_c$ and $V(z)$ are continuous. In practice, we adopt a smoothed version of the step function reconstruction, so that even $\dot{\rho}_c$ and $\dot{V}$ are continuous, see Section \ref{subsec:binned}.}

We focus on two main cases: the first is based on a single redshift bin, the second on four. Thus, in the first case  we consider a single step function, with  a fixed constant value $q_{\rm V}$ from $z=0$  to a transition redshift $z_{\rm trans}$, after which  $q_{\rm V}=0$, the coupling  vanishes and $V$ is constant at higher redshifts. We will elaborate on four variants of this single step function reconstruction scenario in Section \ref{sec:results}, discussing two cases where $z_{\rm trans}$ is kept fixed, a case where we sample over $z_{\rm trans}$ and a case where we assume $V=0$ for $z > z_{\rm trans}$. 

Finally, going beyond the single step function reconstruction, we  want to account for a dynamical  interaction \qv\ with no {\it a priori} assumption of any specific model for its time evolution: to this end, we consider a binned reconstruction of the function \qv, based on several step functions.

It is worth stressing here that $z_{\rm trans}$ is a purely phenomenological parameter, used to implement the step function reconstruction. A true physical model producing an interaction between dark components might indeed imply that such a coupling is active throughout the whole history of the Universe, which would effectively correspond to $z_{\rm trans}=\infty$. However, given our choice of $Q\propto V(z)$, even if the coupling is active at all times it will be effectively vanishing when the vacuum energy becomes negligible. Choosing a $z_{\rm trans}$ corresponding to an era where $V(z)<<\rho_c(z)$ therefore mimics a model in which the coupling is always active and also allows us save computational time, as it only requires solving the differential equations presented in Section \ref{sec:theory} up to $z_{\rm trans}$ (see Section \ref{subsec:const}). 

At the same time, the physical model might imply that the coupling only becomes active when certain conditions are satisfied. Having a low $z_{\rm trans}$ can in principle phenomenologically mimic such a model and obtaining the value of $z_{\rm trans}$ that is preferred by the data would allow us to understand if models with a coupling that is not active at all times are preferred with respect to those in which the transfer of energy between the components is always active.

In the next three subsections, we describe the three main physical scenarios and their implementation through a step function reconstruction; namely a constant $q_{\rm V}$ up to $z_{\rm trans}$ followed by $V=\text{constant}$, a varying $q_{\rm V}(z) $ represented by multiple bins and in which $V=\text{constant}$ after the final bin and finally a constant $q_{\rm V}$ up to the transition redshift $z_{\rm trans}$, after which $V=0$. We then illustrate the effect of the coupling on the cosmological evolution.

\subsection{Constant $q_{\rm V}$ interaction}\label{subsec:const}
With the reconstruction of \qv\ in mind, we elaborate on the five different possibilities, all based on assuming that in some redshift range \qv\ is constant in time, i.e.\ $q_{\rm V}(z)=q_{\rm V}$. Then, in each bin the interaction between dark matter and vacuum energy scales with redshift as $Q(z)\propto H(z)V(z)$. Such an interaction is a sub-case of the linear couplings considered by \cite{Quercellini2008}, and it  greatly simplifies the solutions for $\rho_c$ and $V$, which can be now obtained analytically from equations \eqref{eq:qcdm}-\eqref{eq:qvac}. 

Setting initial conditions at $z=0$ gives
\begin{equation}
\label{eq:rhom}
\rho_c(z)=\rho_c^0a^{-3}+V_0 \frac{q_{\rm V}}{q_{\rm V}-3}\left(a^{-3}-a^{-q_{\rm V}}\right),
\end{equation}
\begin{equation}\label{eq:V}
V(z)=V_0a^{-q_{\rm V}},
\end{equation}
where $\rho_{c}^0$ and $V_0$ are the present values of the energy density of CDM and vacuum, respectively. Furthermore, the equations for matter perturbations $\delta_c$ follow Eq.~(\ref{eq:densitycontrast}). 

Analytical expressions similar to \eqref{eq:rhom}-\eqref{eq:V} can be found in different redshift bins, in a way that guarantees the continuity of $\rho_c$ and $V$ across bin boundaries.

It is worth noticing at this point that the choice of a constant \qv \ is a strong assumption that has to be taken with a pinch of salt: it conveniently simplifies the equations but can give an unphysical model\footnote{For instance,  in an  over-simplified model based on  a negative constant $q_{\rm V}$ at all times the cold dark matter density $\rho_c$ would become negative at some point.}; we use  it here only to give a  phenomenological representation of a generic interaction in various redshift ranges, up to $z=0$.  

Hence, a first step we can take towards a more general description of the coupling is to consider a single step function reconstruction for  \qv, i.e.\ a \qv\ that remains constant up to a certain redshift $z_{\rm trans}$ and vanishes for higher redshifts; this corresponds to a cosmology equivalent to $\Lambda$CDM in the distant past, undergoing a transition at $z_{\rm trans}$ where the coupling is turned on and densities and perturbations start to scale as in the constant $q_{\rm V}$ case. 

\subsection{Binned reconstruction}\label{subsec:binned}
In order to allow for a variation in redshift of the coupling function \qv, we reconstruct its evolution using a number of redshift bins $N$, with the $i^{\rm th}$ bin being enclosed in the range $[z_{i-1},z_i]$, with $z_0=0$ and $i=1,...,N$. For each of these bins the value at the centre of the range ($\bar{z}_i$) is $q_i = q_{\rm V}(\bar{z}_i)$ and we assume the function to take this constant value within the entire redshift bin. With this choice, we can generally reconstruct the value of the function at any point as 
\begin{align}
\label{eq:sharpbin}
q_{\rm V}(z)=q_1+\sum_{i=1}^{N-1}\left(q_{i+1}-q_{1}\right)\left[\theta_H(z-z_i)-\theta_H(z-z_{i+1})\right]
\intertext{or, equivalently,}
\label{eq:sharpbin1}
q_{\rm V}(z)=q_1+\sum_{i=1}^{N-1}\left(q_{i+1}-q_{i}\right)\left[\theta_H(z-z_i)\right]\,
\end{align}
\noindent
where $\theta_H$ is the Heaviside function. We choose however to adjust this reconstruction by introducing a smoothing at the border of the bins,  controlled by the parameter $s$, substituting the Heaviside functions with smooth steps based on hyperbolic tangent functions. This allows us to avoid sharp transitions between values of the function \qv, which could lead to numerical problems. Given that no derivatives of the coupling enter our equations, this should not be an issue in our case, but even so, we rewrite the reconstructed function as

\begin{equation}\label{eq:smoothbin}
q_{\rm V}(z)=q_1+\sum_{i=1}^{N-1}\frac{q_{i+1}-q_{i}}{2}\left[1+\tanh{\left(s\frac{z-z_i}{z_i-z_{i-1}}\right)}\right].
\end{equation}
Using \eqref{eq:sharpbin1} in equations \eqref{eq:qcdm}-\eqref{eq:qvac} gives analytic expressions similar to \eqref{eq:rhom}-\eqref{eq:V} in each bin, matched at the bin boundaries; using \eqref{eq:smoothbin} gives a smoothed version of the same \qv. With this, we numerically obtain the densities $\rho_c$ and $V$ such that their derivatives $\dot{\rho}_c$ and $\dot{V}$ are continuous through the bin boundaries. We have checked that the numerical and analytical solutions for $\rho_c$ and $V$ match extremely well. 

\subsection{Seeded vacuum energy}
In the cosmology described above, there is a standard $\Lambda$CDM evolution at high redshifts until the coupling switches on at $z_{\rm trans}$ and the vacuum and CDM energies can begin to interact. Instead, in the \textit{seeded vacuum energy} case, or \textit{SVE}, we have designed a reconstruction that mimics a physical scenario in which for $z >z_{\rm trans}$ we have a pure CDM (Einstein -- de Sitter) evolution, rather than  $\Lambda$CDM. In this scenario the coupling causes the vacuum energy to suddenly grow from zero up to a  `seed' value at $z_{\rm trans}$, a kind of fast transition; cf.\ \cite{Piattella:2009kt,Bertacca:2010mt} for a similar idea for unified dark matter models. Then, at lower redshifts, the interaction is characterized  as in the previous  cases, i.e.\ with a constant $q_{\rm V}$, and the vacuum evolves accordingly. The free parameter, $z_{\rm trans}$ allows this rapid growth of vacuum to a non-zero value to occur even at very late times. 

In practice, this set-up is achieved by some reverse engineering in \texttt{CAMB}. Since the coupling function $Q$ is proportional to $V$, if $V$ remained practically zero for the entire cosmic history we would never have any interaction. Instead, we `seed' the growth of vacuum by inducing a sudden spike in its density at $z_{\rm trans}$. The vacuum energy $V$ can then grow to a finite value and the transfer of energy between the vacuum and CDM via the coupling can begin.

\subsection{Effects of the coupling}
\begin{figure}
\includegraphics[width=\columnwidth]{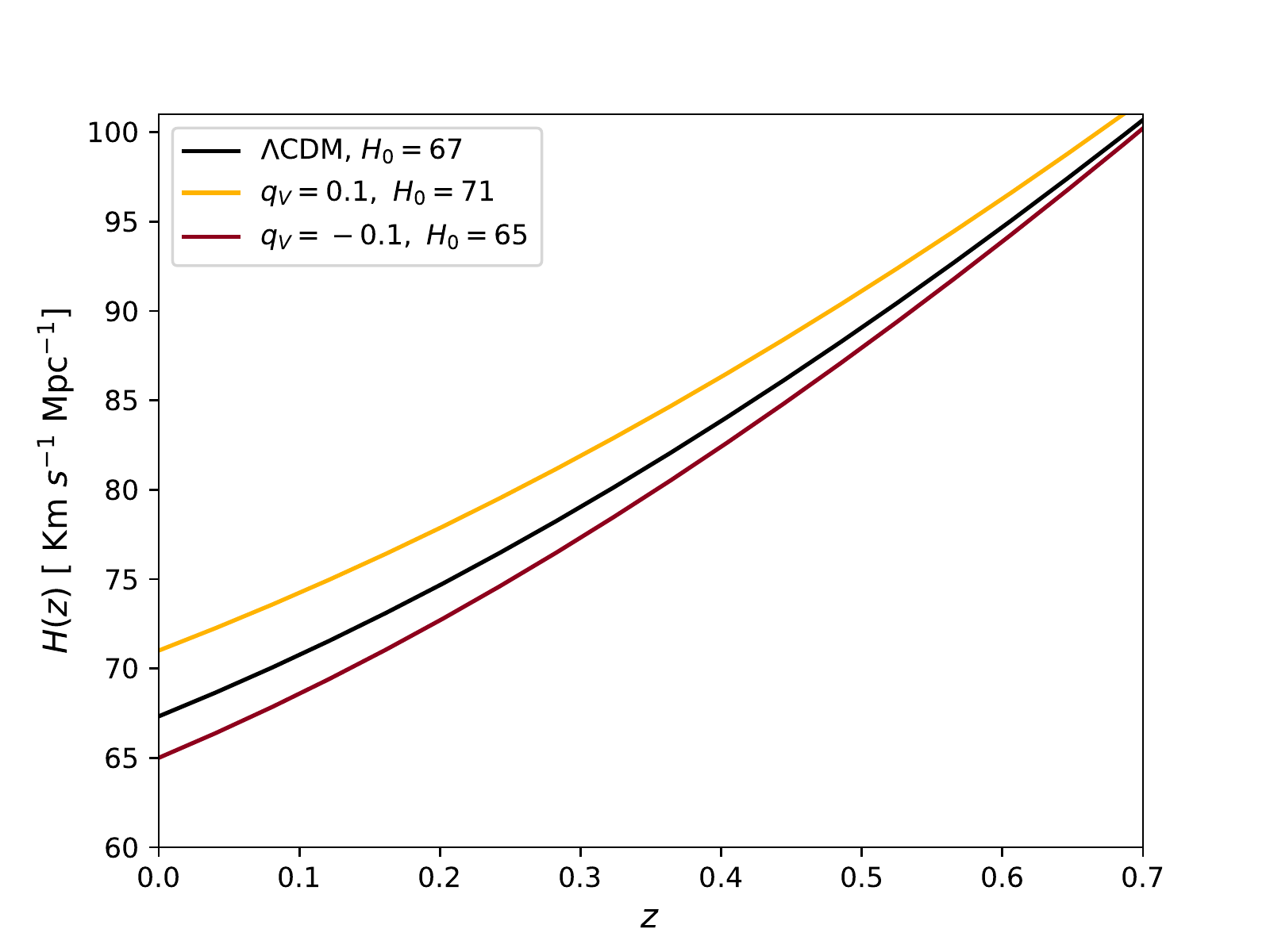}
\caption{The evolution of the Hubble function $H(z)$ for 3 cosmologies resulting in the same angular size of the sound horizon at recombination. Except for $q_{\rm V}$ and $H_0$, whose values are shown in the label, all the other primary parameters are fixed to the Planck 2015 best fit.\label{fig:hubble}}
\end{figure}

As mentioned in the Introduction, we are interested in the ability of these models to ease the tensions between low and high redshift observations. In particular we focus on the tension between the local determination of $H_0$ and that inferred from CMB measurements of the angular size of the sound horizon at recombination, $\theta_\mathrm{MC}$. In Figure \ref{fig:hubble} we show the $H(z)$ obtained for 3 different values of $q_{\rm V}$ and the same value of $\theta_\mathrm{MC}$, also highlighting the resulting value of $H_0$, while the other cosmological parameters, i.e. the densities $\Omega_bh^2$ and $\Omega_ch^2$, primordial power spectrum amplitude and tilt $A_s$ and $n_s$ and the optical depth $\tau$, are fixed to the best fit of Planck 2015 \cite{Ade:2015xua}. We find that starting from the Planck value of $\theta_\mathrm{MC}$, a positive $q_{\rm V}$ leads to higher values of $H_0$ with respect to $\Lambda$CDM, thus moving in the direction required to ease the tension. 

\begin{figure}
\includegraphics[width=\columnwidth]{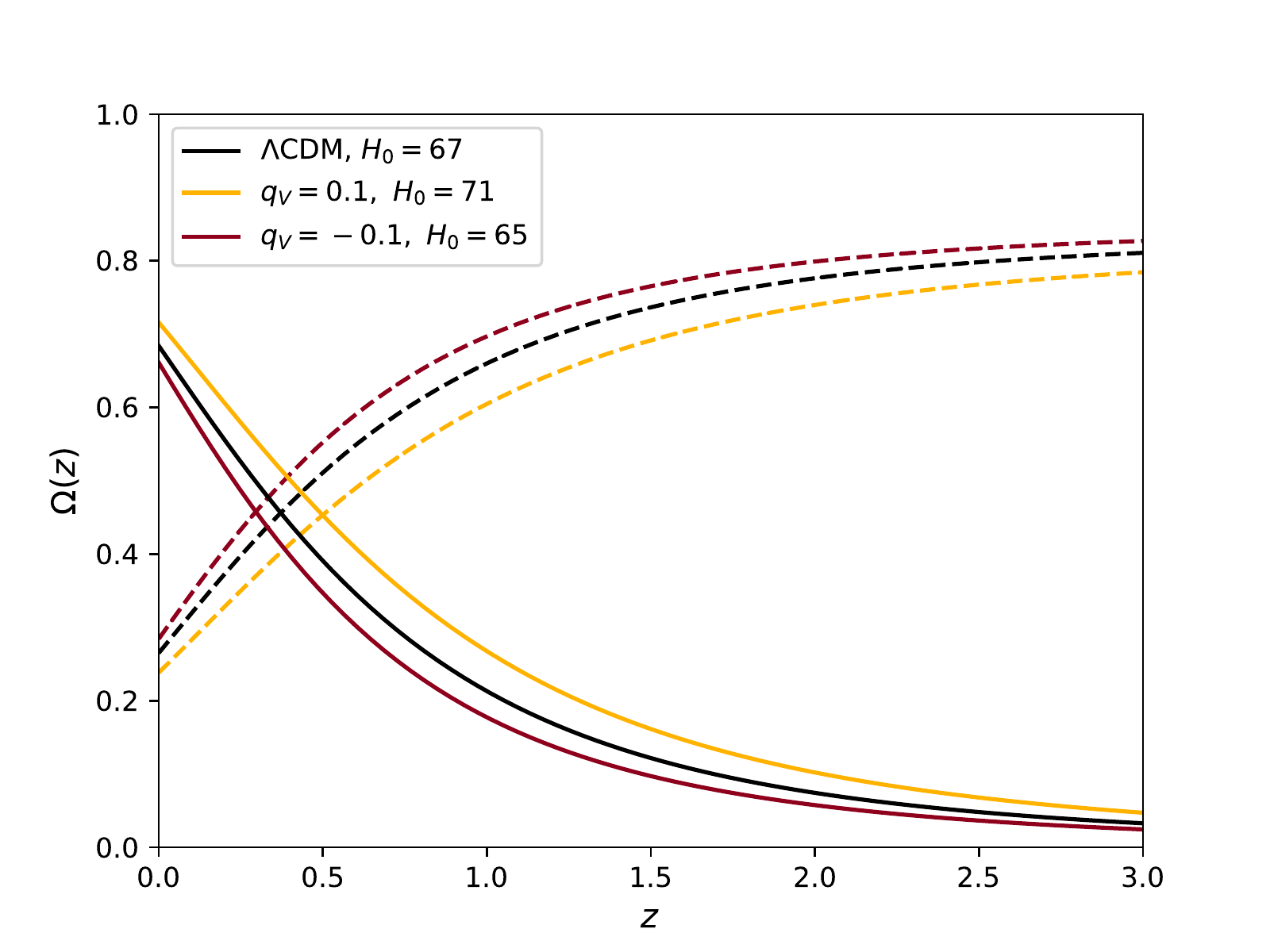}
\caption{The evolution of the matter (dashed lines) and vacuum density (solid lines) parameters as a function of redshift, for a small positive and negative coupling. The \LCDM~case is shown in blue. Except for $q_{\rm V}$ and $H_0$, whose values are shown in the label, all the other primary parameters are fixed to the Planck 2015 best fit.\label{fig:omegas}}
\end{figure}

\begin{figure}
\includegraphics[width=\columnwidth]{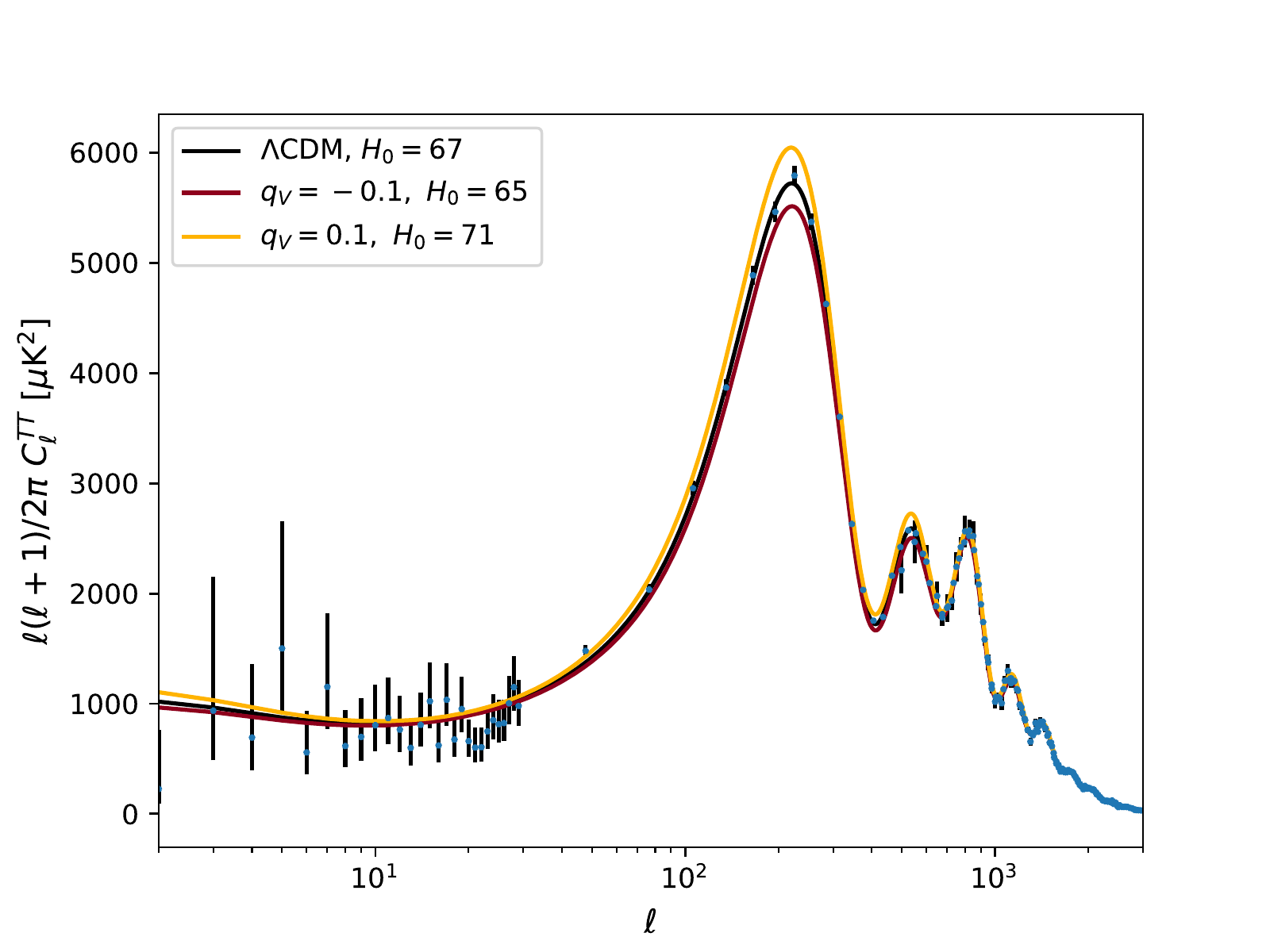}
\caption{The CMB TT power spectrum for 3 cosmologies resulting in the same angular size of the sound horizon at recombination. Except for $q_{\rm V}$ and $H_0$, whose values are shown in the label, all the other primary parameters are fixed to the Planck 2015 best fit. The data points are the TT observations of Planck 2015.\label{fig:cls}}
\end{figure}

\begin{figure}
\includegraphics[width=\columnwidth]{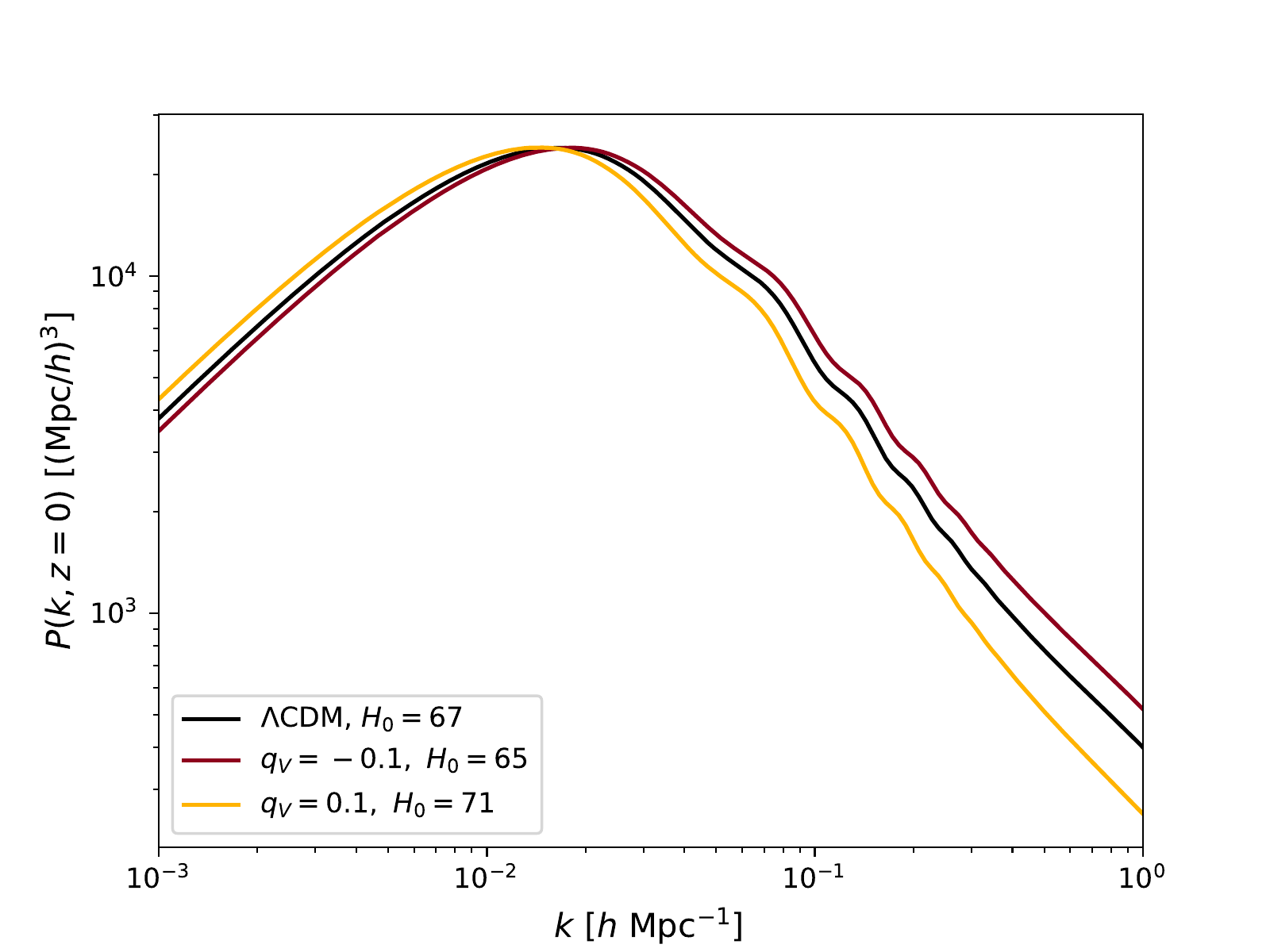}
\caption{The matter power spectrum at $z=0$ for 3 cosmologies resulting in the same angular size of the sound horizon at recombination. Except for $q_{\rm V}$ and $H_0$, whose values are shown in the label, all the other primary parameters are fixed to the Planck 2015 best fit. The \LCDM~case is plotted in blue. \label{fig:matterpower}}
\end{figure}

Figures \ref{fig:omegas},\ref{fig:cls} and \ref{fig:matterpower} illustrate different aspects of the same three cosmologies. Given the definition of $Q$ in Eq.~(\ref{eq:bigQ}), a negative value for $q_{\rm V}$ implies that cold dark matter is decaying into the vacuum, thus with the values of the density parameters $\Omega_ch^2$ and $\Omega_bh^2$ fixed at $z=0$ we end up with a higher matter density in the past (see Figure \ref{fig:omegas}). However, because the cosmologies shown here have the same present value of the matter density $\Omega_ch^2$, they will have significantly different matter abundances at early times; this impacts other observables, e.g. CMB power spectra which are significantly affected by the amount of matter (see Figure \ref{fig:cls}). Therefore if the only free parameters considered  are $q_{\rm V}$ and $H_0$ one would expect a positive correlation between the two, but it is crucial not to neglect the effect of matter abundance on predictions for cosmological probes and the resulting degeneracy of $\Omega_ch^2$ with $q_{\rm V}$ and $H_0$.

In Figure \ref{fig:matterpower} the effect of the coupling on the evolution of perturbations is shown through its effect on the matter power spectrum $P(k,z)$; we can see that a positive value of $q_{\rm V}$ suppresses the amplitude of $P(k,z)$, while on the contrary this is increased by a negative $q_{\rm V}$. We stress that even though the results we comment on here refer to a case with constant $q_{\rm V}$ up to $z=1$ and vanishing at higher redshifts, the same qualitative behaviour also holds for different choices of the redshift evolution of $q_{\rm V}$.

\section{Data and analysis method}\label{sec:analysis}
We want to compare the predictions of the interacting vacuum scenario with recent cosmological data. For this analysis we consider the Planck 2015 measurements of the CMB temperature and polarization~\citep{Aghanim:2015xee,Ade:2015xua}. For the Planck likelihood, we also vary the nuisance parameters that are used to model foregrounds as well as instrumental and beam uncertainties. We note that at the time of writing, the new Planck 2018 likelihood was not publicly available, but given the similarities between the Planck 2015 and 2018 results we do not expect that our results would change significantly were we to use the 2018 data presented in \cite{Aghanim:2018eyx}.

In addition to the Planck CMB data, we utilize the BAO measurement from the 6dF Galaxy Survey \citep{Beutler2011}, the BAO scale measurement from the SDSS DR7 Main Galaxy Sample \citep{Ross2015} and the combined  BAO and RSD data from the SDSS DR12 consensus release \citep{Alam:2016hwk} (data points listed in Table \ref{tab:lowzdata}), together with the JLA Type Ia supernovae sample \citep{Betoule:2014frx}. We refer to the combined datasets as Planck + Low-z, with Low-z referring to the combination of all dataset at redshifts lower than recombination.

\begin{table}
\centering
\begin{tabular}{|c|c|c|c|}
\hline
Quantity  & $z_{\text{eff}}$ & Measurement  & Source \\
\hline
$D_v$ & 0.106 & $457 \pm 27 (r_s / r_{s, \text{fid}})$ Mpc & \cite{Beutler2011} \\
$D_v$ & 0.15 & $ (664 \pm 25) (r_s / r_{s, \text{fid}})$ Mpc & \cite{Ross2015}\\
$D_v$ & 0.32 & $ (1270 \pm 14) (r_s / r_{s, \text{fid}})$ Mpc &\cite{Alam:2016hwk}\\
$D_v$ & 0.57 & $ (2033 \pm 21) (r_s / r_{s, \text{fid}})$ Mpc &\cite{Alam:2016hwk}\\
$f\sigma_8$ & 0.32 & 0.392 &\cite{Alam:2016hwk}\\
$f\sigma_8$ & 0.57 & 0.445 &\cite{Alam:2016hwk}\\
\hline
\hline
\end{tabular}
\caption{This table lists the BAO and $f \sigma_8$ datapoints used in our analysis. The parameter $D_v$ is a distance scale, defined as $D_v(z)= \left[(1+z)^2 D_A^2(z) \frac{cz}{H_0 E(z)}\right]^{1/3}$, $D_A$ being the angular diameter distance \citep{Beutler2011}, and $f \sigma_8$ is the value of the linear growth rate $f$ multiplied by $\sigma_8$, the amplitude of the linear matter power spectrum on a scale of $8 h^{-1}$ Mpc.}\label{tab:lowzdata}
\end{table}

\subsection{Implementation in \texttt{CAMB}}\label{subsec:implementation}
Now that we have chosen our methods of reconstruction, we need to obtain predictions for the cosmological observables. In order to do so we use the Einstein-Boltzmann Code for the Anisotropies in the Microwave Background (\texttt{CAMB}); we modify the code so that it uses the $\rho_c(z)$ and $V(z)$ of our model rather than those computed internally within the $\Lambda$CDM framework. We therefore add a new module which solves the differential equations \eqref{eq:qcdm} and \eqref{eq:qvac}, with \qv~ computed at each redshift according to the methods described in Section \ref{sec:qreconst}. We use a Runge--Kutta algorithm, starting from the present day with initial conditions 
\begin{eqnarray}\label{eq:inicon}
\rho_c^0=3H_0^2\Omega_c, \nonumber \\
V_0=3H_0^2\Omega_\Lambda, 
\end{eqnarray}
and then evolving the equations backwards in time. To solve the equations for CDM perturbations we make use of the routines present in \texttt{CAMB}, modifying the equation for CDM with the extra source term proportional to \qv~ described in \eqref{eq:densitycontrast}.

On top of this, we make use of the MCMC sampler \texttt{CosmoMC} \citep{Lewis:2002ah, Lewis:2013hha} to sample the parameter space and compare our predictions with the cosmological data mentioned above. The six sampled parameters are therefore those of the minimal $\Lambda$CDM: the baryon and cold dark matter densities at present day, $\Omega_b h^2$ and $\Omega_ch^2$; the optical depth, $\tau$; the primordial power spectrum amplitude and tilt, $A_s$ and $n_s$, and the Hubble constant $H_0$. 

Furthermore, we also consider additional parameters depending on the specific cosmology we investigate:
\begin{itemize}
\item {\it Cfix}: the constant coupling $q_{\rm V}$ with uniform prior $[-6,3]$, controlling the evolution of the densities up to a fixed $z_{\rm trans}=3000$, with standard $\Lambda$CDM evolution at higher redshifts. We also consider a variation on this in which $z_{\rm trans} = 0.9$, to compare directly with \cite{Salvatelli2014}.
\item {\it Cvar}: the constant coupling $q_{\rm V}$ and the varying $z_{\rm trans}$ with uniform priors $[-6,3]$ and $[0.1,10]$ respectively. At redshifts higher than $z_{\rm trans}$ the coupling is turned off and we then have standard $\Lambda$CDM evolution. In order to test the stability of the results changing the prior choice, we also explored a logarithmic prior on $z_{\rm trans}$, including also higher values of this parameter, finding no significant differences in our results. We choose therefore to present in the paper only the results obtained with the uniform prior.
\item \textit{SVE}: a constant $q_{\rm V}$ and the varying transition redshift $z_{\rm trans}$. At redshifts higher than the transition redshift, CDM evolves in the standard way while $V(z)$ smoothly transitions to zero from its value at $z_{\rm trans}$ according to the solution of the differential equations. For these parameters we also use the uniform priors $[-6,3]$ and $[0.1,10]$ respectively.
\item {\it 4bins}: $N=4$ low redshift bins $q_i$, with uniform priors $[-6,3]$, used to reconstruct the evolution in time of the coupling function \qv, with a return to standard $\Lambda$CDM for redshifts higher than the last bin. The number and redshift of the considered bins ($z_i \in \{0.3, 0.9, 2.5, 10\}$) are chosen in order to compare our results with that from previous work by \cite{Salvatelli2014}.
\end{itemize}

\noindent
The choice of the prior range $[-6,3]$ for the $q_{\rm V}$ parameters arises from the fact that $\rho_c$ in  \eqref{eq:rhom} becomes singular when $q_{\rm V}=3$. While higher values of the coupling are theoretically possible, we choose to limit the parameter space to the non-pathological part, in order to avoid issues with the sampling. Indeed, we find that this prior is sufficiently broad as to have no effect on our results. 

A summary of the priors used on all parameters can be found in Table \ref{tab:priors}.

\begin{table}
\centering
\begin{tabular}{|c|c|}
\hline
Parameter          & Prior range \\
\hline
$\Omega_bh^2$      &  $[0.005,0.1]$\\
$\Omega_ch^2$      &  $[0.001,0.99]$ \\
$H_0$              &  $[50,100]$\\
$\tau$             &  $[0.01,0.8]$\\
$\log{10^{10}A_s}$ &  $[2.0,4.0]$\\
$n_s$              &  $[0.8,1.2]$\\
$q^i_{\rm V}$      &  $[-6,3]$  \\
$z_{\rm trans}$    &  $[0.1,10]$ \\
\hline
\hline
\end{tabular}
\caption{Prior ranges on the cosmological parameters sampled in our analysis. The prior range on $z_{\rm trans}$ refers to the {\it Cvar} and {\it SVE} cases, while in the rest of the analysis this parameter is fixed.}\label{tab:priors}
\end{table}

\section{Results} \label{sec:results}
In this section we present the results of our investigation, beginning with the two {\it Cfix} cases where the interaction is characterised by a  constant  parameter $q_{\rm V}$  up to a transition redshift, moving to the cases where the transition redshift $z_{\rm trans}$ is allowed to vary ({\it Cvar} and {\it SVE}) and finally the {\it 4bins} case. We remark again that any integration is performed with initial values set today at $z=0$. In particular a non-zero value for the vacuum $V_0$ is set as in \eqref{eq:inicon}. 

In Table \ref{tab:constres} we summarise results for the five cases; we report the marginalized constraints on the primary parameters sampled in our analysis, adding also the combination of derived parameters $\sigma_8\Omega_m^{1/2}$, useful to assess the status of the tensions between high and low redshift probes.

\subsection{{\it Cfix} case}
As a baseline result, we report the constraints obtained assuming a constant value $q_{\rm V}$ for the coupling, up to a fixed redshift $z_{\rm trans}=3000$. At higher redshifts, the interaction is turned off ($q_{\rm V}(z>z_{\rm trans})=0$) and the vacuum assumes a constant value $V=V(z=z_{\rm trans})$.  This choice is made so that the interaction affects the evolution of CDM and vacuum only after the last scattering surface; however, given our choice of $Q\propto V$, the interaction is negligible during the matter dominated era.

\begin{figure}
\includegraphics[width=.5\textwidth]{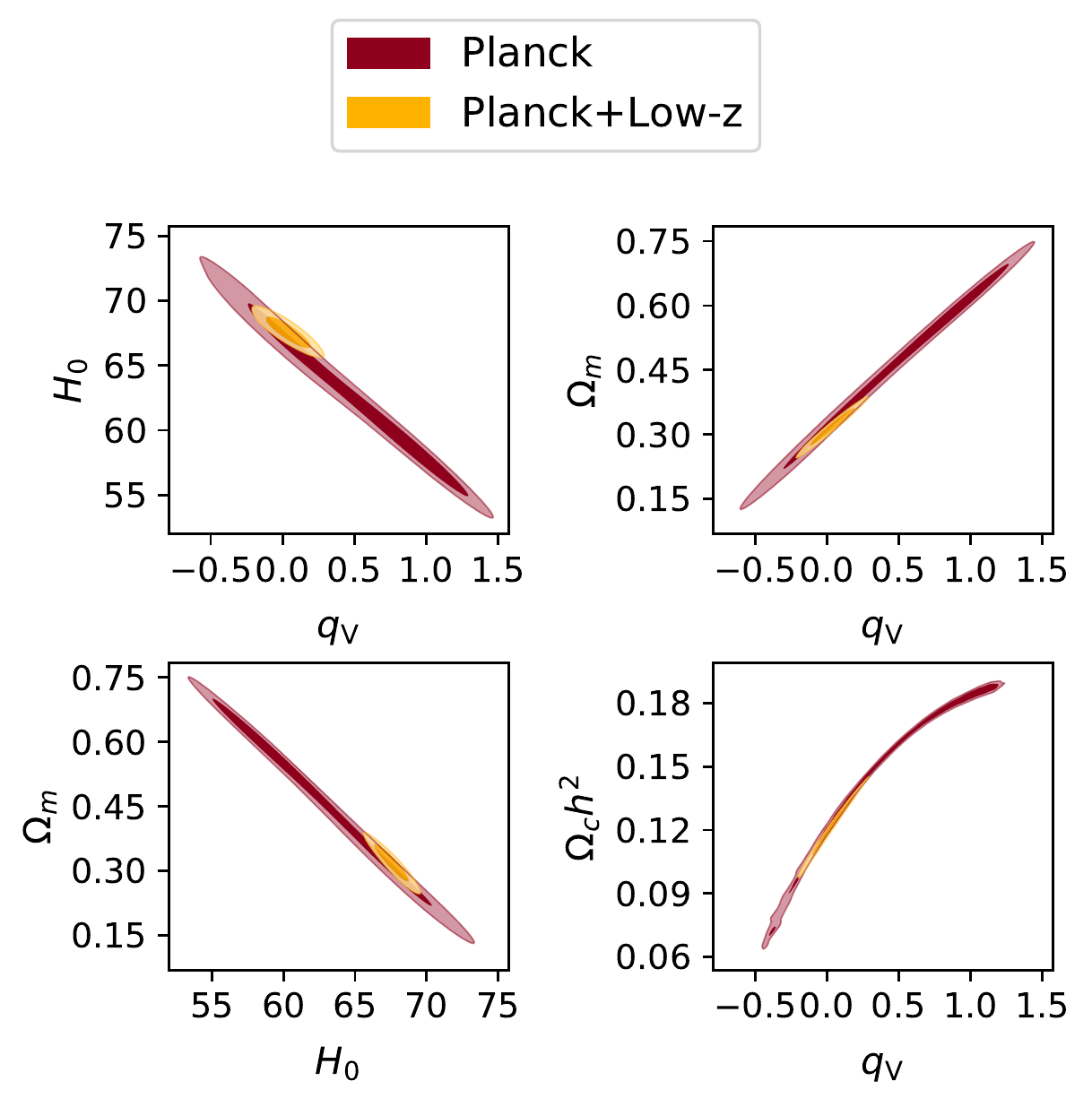}
\caption{{\it Cfix} case with $z_{\rm trans} = 3000$: $68$\% and the $95$\% confidence level marginalized contours on $H_0$, $q_{\rm V}=q_{\rm V}(z\leq 3000)$ and $\Omega_m$ as obtained in the analysis with the Planck (red) and Planck + Low-z (yellow) datasets.\label{fig:triconst}}
\end{figure}

 In Figure \ref{fig:triconst} we show the 2D joint marginalized contours of $q_{\rm V}$ with $H_0$, $\Omega_m$ and $\Omega_c h^2$.
We point out that the constraints placed by Planck on $q_{\rm V}$ and $H_0$ are strongly degenerate. This effect is due to the change in the Universe's expansion history caused by the interaction: we find that a larger $H_0$ requires a smaller coupling parameter $q_{\rm V}$ in order to recover the same expansion history. A similar degeneracy is also present between $q_{\rm V}$ and $\Omega_m$. In general, the CMB data prefer positive values of $q_{\rm V}$. Negative values of $q_{\rm V}$ imply that we would have a smaller CDM density at late times (see bottom right panel of Figure \ref{fig:triconst}), which would boost the amplitude of the acoustic peaks in the CMB temperature--temperature power spectrum by such an amount that the change could not be compensated for by equivalent changes in the other cosmological parameters.

We find that the Planck data alone allow for the coupling $q_{\rm V}$ to be non-vanishing; however, the $\Lambda$CDM limit of this model is within the 68\% confidence level region. The degeneracies between $q_{\rm V}$, $H_0$ and $\Omega_m$ are broken when the Low-z datasets are added to Planck. This is because the data directly probe the redshift range where the interaction is primarily effective. The combination of the Planck and Low-z data does not allow $q_{\rm V}$ to greatly deviate  from zero and the cosmology is therefore very similar to $\Lambda$CDM. 

\begin{table*}
\centering
\begin{tabular}{|c|c|c|c|}
\hline
    Parameter  & Case & Planck & Planck + Low-z\\
\hline
\hline
 & \textit{Cfix} & $0.02226\pm 0.00022$ & $0.02235\pm 0.00015$  \\
 & \textit{Cfix ($z_{\rm trans} = 0.9$)} & $0.02226^{+0.00014}_{-0.00020}$ & $0.02235\pm 0.00014$\\
$\Omega_b h^2$ & \textit{Cvar} & $0.02222\pm 0.00015 $ & $0.02234\pm 0.00014$  \\
 & \textit{SVE} & $0.02224\pm 0.00016$ & $0.02235\pm 0.00015$  \\
 & \textit{4bins} & $0.02224\pm 0.00015$ & $0.02226\pm 0.00016$  \\
\hline
 & \textit{Cfix} & $0.131\pm 0.040 $ & $0.122^{+0.011}_{-0.0089}$  \\
 & \textit{Cfix ($z_{\rm trans} = 0.9$)} & $0.118^{+0.025}_{-0.038}$ & $0.130\pm 0.015$\\
$\Omega_c h^2$ & \textit{Cvar} & $0.153^{+0.047}_{-0.031}$ & $0.124\pm 0.012$  \\
 & \textit{SVE} & $0.150^{+0.049}_{-0.024}$ & $0.124\pm 0.011$  \\
 & \textit{4bins} & $0.132^{+0.031}_{-0.056}$ & $0.117^{+0.020}_{-0.045}$  \\
\hline
 & \textit{Cfix} & $0.080^{+0.021}_{-0.017}$ & $0.077\pm 0.017$  \\
 & \textit{Cfix ($z_{\rm trans} = 0.9$)} & $0.080^{+0.018}_{-0.015}$ & $0.078\pm 0.016$\\
$\tau$ & \textit{Cvar} & $0.080\pm 0.017$ & $0.077\pm 0.016$  \\
 & \textit{SVE} & $0.079\pm 0.016$ & $0.076\pm 0.017$  \\
 & \textit{4bins} & $0.081\pm 0.017$ & $0.074\pm 0.017$  \\
\hline
 & \textit{Cfix} & $3.094^{+0.039}_{-0.032} $ & $3.084\pm 0.033 $  \\
 & \textit{Cfix ($z_{\rm trans} = 0.9$)} & $3.094^{+0.029}_{-0.033}$ & $3.087\pm 0.032$\\
$\log{10^{10}A_s}$ & \textit{Cvar} & $3.094\pm 0.034$ & $3.084\pm 0.031$  \\
 & \textit{SVE} & $3.093\pm 0.032$ & $3.084\pm 0.033$  \\
 & \textit{4bins} & $3.098\pm 0.032$ & $3.082\pm 0.034$  \\
\hline
 & \textit{Cfix} & $0.9647^{+0.0048}_{-0.0062}$ & $0.9681\pm 0.0043$  \\
 & \textit{Cfix ($z_{\rm trans} = 0.9$)} & $0.9658^{+0.0042}_{-0.0062}$ & $0.9684\pm 0.0040$\\
$n_s$ & \textit{Cvar} & $0.9643\pm 0.0047$ & $0.9679\pm 0.0041$  \\
 & \textit{SVE} & $0.9646\pm 0.0048$ & $0.9682\pm 0.0043$  \\
 & \textit{4bins} & $0.9644\pm 0.0045$ & $0.9655\pm 0.0047$  \\
\hline
 & \textit{Cfix} & $62.3^{+3.2}_{-6.2}$ & $67.54\pm 0.80$  \\
 & \textit{Cfix ($z_{\rm trans} = 0.9$)} & $67.05 \pm 2.1$ & $67.26 \pm 0.86$\\
$H_0$ & \textit{Cvar} & $62.2^{+4.9}_{-5.5}$ & $67.50\pm 0.81$  \\
 & \textit{SVE} & $61.9\pm 5.2$ & $67.46\pm 0.86$  \\
 & \textit{4bins} & $64.0\pm 4.8$ & $67.33\pm 0.80$  \\
\hline
&  \textit{Cfix} & $0.4652^{+0.0075}_{-0.022}$ & $0.452^{+0.011}_{-0.014}$  \\
& \textit{Cfix ($z_{\rm trans} = 0.9$)} & $0.4752 \pm 0.037$& $0.446 \pm 0.017$\\
 $\sigma_8\Omega_m^{1/2}$  &  \textit{Cvar} & $0.4614^{+0.0088}_{-0.021}$ & $0.451^{+0.012}_{-0.015}$ \\
& \textit{SVE} & $0.461^{+0.012}_{-0.025}$ & $0.450\pm 0.016$  \\
& \textit{4bins} & $0.481^{+0.064}_{-0.076}$ & $0.482\pm 0.055$  \\
\hline
 &  \textit{Cfix} & $0.52^{+0.65}_{-0.77}$ & $0.04\pm 0.10$  \\
 & \textit{Cfix ($z_{\rm trans} = 0.9$)} & $0.059 \pm 0.39$&  $0.14 \pm 0.19$\\
$q_{\rm V}$ &  \textit{Cvar} & $0.59\pm 0.53$ & $0.07^{+0.11}_{-0.14}$ \\
& \textit{SVE} & $0.62\pm 0.60$ & $0.06\pm 0.12$  \\
\hline
$ q_{1}$  & \textit{4bins} & $0.0^{+1.2}_{-1.5}$ & $-0.42^{+0.51}_{-1.0}$  \\
$ q_{2}$  & \textit{4bins} & $0.3^{+1.9}_{-1.2}$ & $0.88^{+0.82}_{-0.66}$  \\
$ q_{3}$  & \textit{4bins} & $>-2.7$ & $-0.62^{+1.3}_{-0.91}$  \\
$ q_{4}$  & \textit{4bins} & unconstrained & unconstrained  \\
\hline
 &  \textit{Cfix} & $-$ & $-$  \\
 & \textit{Cfix ($z_{\rm trans} = 0.9$)} & $-$&  $-$\\
$z_{\rm trans}$ &  \textit{Cvar} & unconstrained & unconstrained \\
& \textit{SVE} & $>1.7$ & $>1.4$  \\
\hline
\hline
\end{tabular}
\caption{Marginalized values of the parameters and their $68\%$ confidence level bounds, obtained using Planck and Planck + Low-z. When only upper or lower bounds are found, we report the $95\%$ confidence level limit.}\label{tab:constres}
\end{table*}

\subsection{\textit{Cfix} with low transition redshift} \label{subsec:Low-z_cfix}
We now consider a \textit{Cfix} case in which we set the transition redshift to $z_{\rm trans} = 0.9$. This allows us to make a direct comparison with the so-called $q_{34}$ case presented in \cite{Salvatelli2014}, in which it was found that a null interaction was excluded at the 99\% confidence level.

This \textit{Cfix} case should be seen as a simple single-step function reconstruction of an interaction that is negligible for $z > z_{\rm trans} = 0.9$. It is a single-parameter reconstruction where, as in \cite{Salvatelli2014} and in comparison to our \textit{4bins} case of section \ref{sec:4bins}, the first two bins are grouped together, with no interaction for  $z> z_{\rm trans} = 0.9$. Note that in \cite{Salvatelli2014} the  $z_{\rm trans} = 0.9$ value was also chosen because it was the best fit value resulting from a two parameter analysis, similar to our {\it Cvar} case in the next section.

\begin{figure}
\includegraphics[width=.5\textwidth]{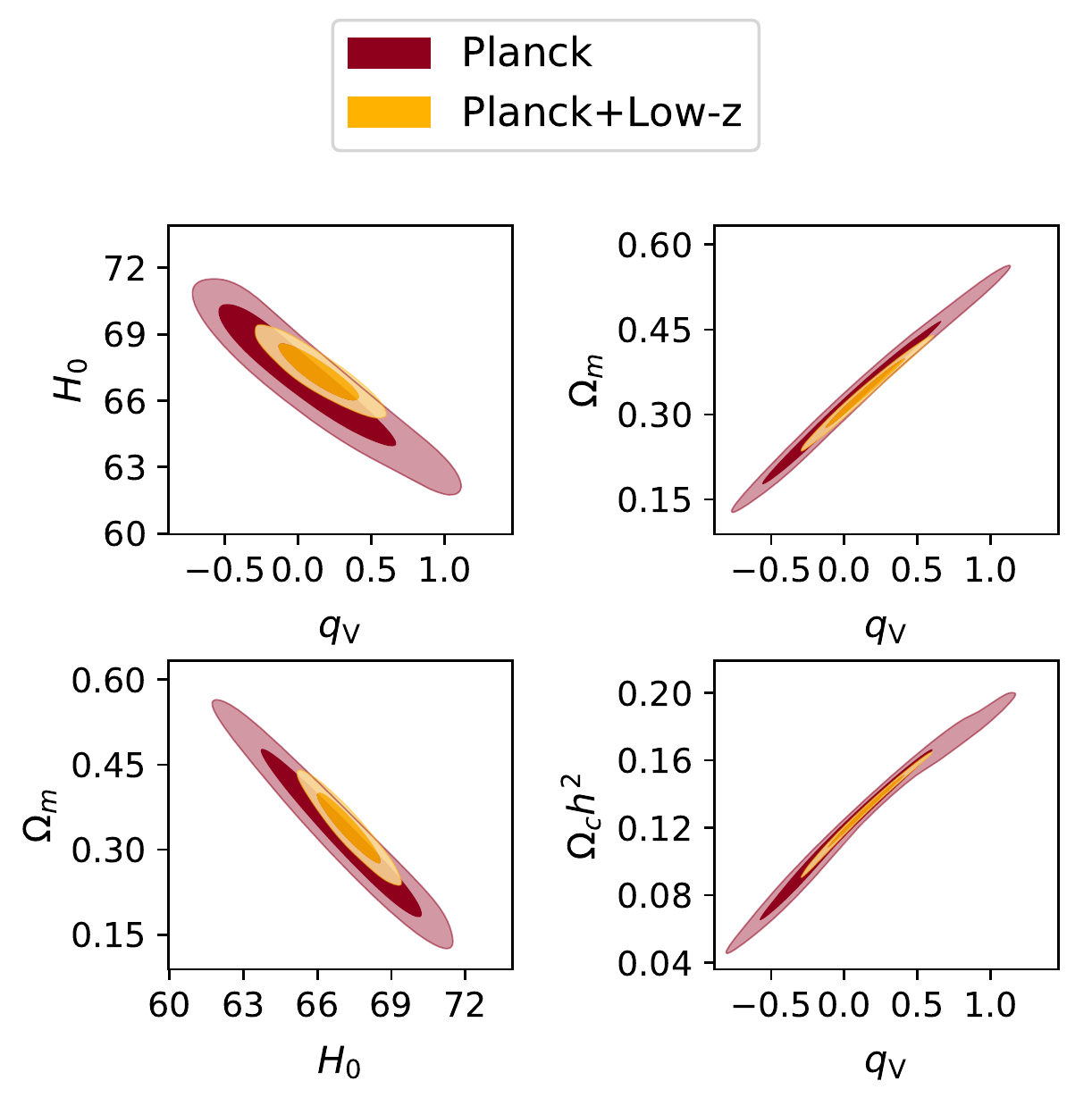}
\caption{{\it Cfix} case with $z_{\rm trans} =0.9$: $68$\% and the $95$\% confidence level marginalized contours on $H_0$, $q_{\rm V}=q_{\rm V}(z\leq 0.9)$, $\Omega_m$ and $\Omega_c h^2$ as obtained in the analysis with the Planck (red) and Planck + Low-z (yellow) datasets.\label{fig:cfixztrans09}}
\end{figure}

Our results for this case are similar to that of the  \textit{Cfix} case with $z_{\rm trans} = 3000$. However, in this case, the CMB bound on $q_{\rm V}$, and consequently the bound on the degenerate cosmological parameters, is less broad and more directly centred on $q_{\rm V} = 0$ with respect to the $z_{\rm trans} = 3000$ case; this is due to the fact that the coupling is active for less time and therefore values of $q_{\rm V}$ that are significantly different from zero cannot be compensated by changes in $\Omega_c h^2$. This result differs from that found by \cite{Salvatelli2014} in that we do not exclude the \LCDM~limit of $q_{\rm V} =0$ at any confidence level. The marginalized 2D joint distributions for the relevant parameters in this case are shown in Figure~\ref{fig:cfixztrans09}.

\subsection{{\it Cvar} case}
\begin{figure}
\includegraphics[width=.5\textwidth]{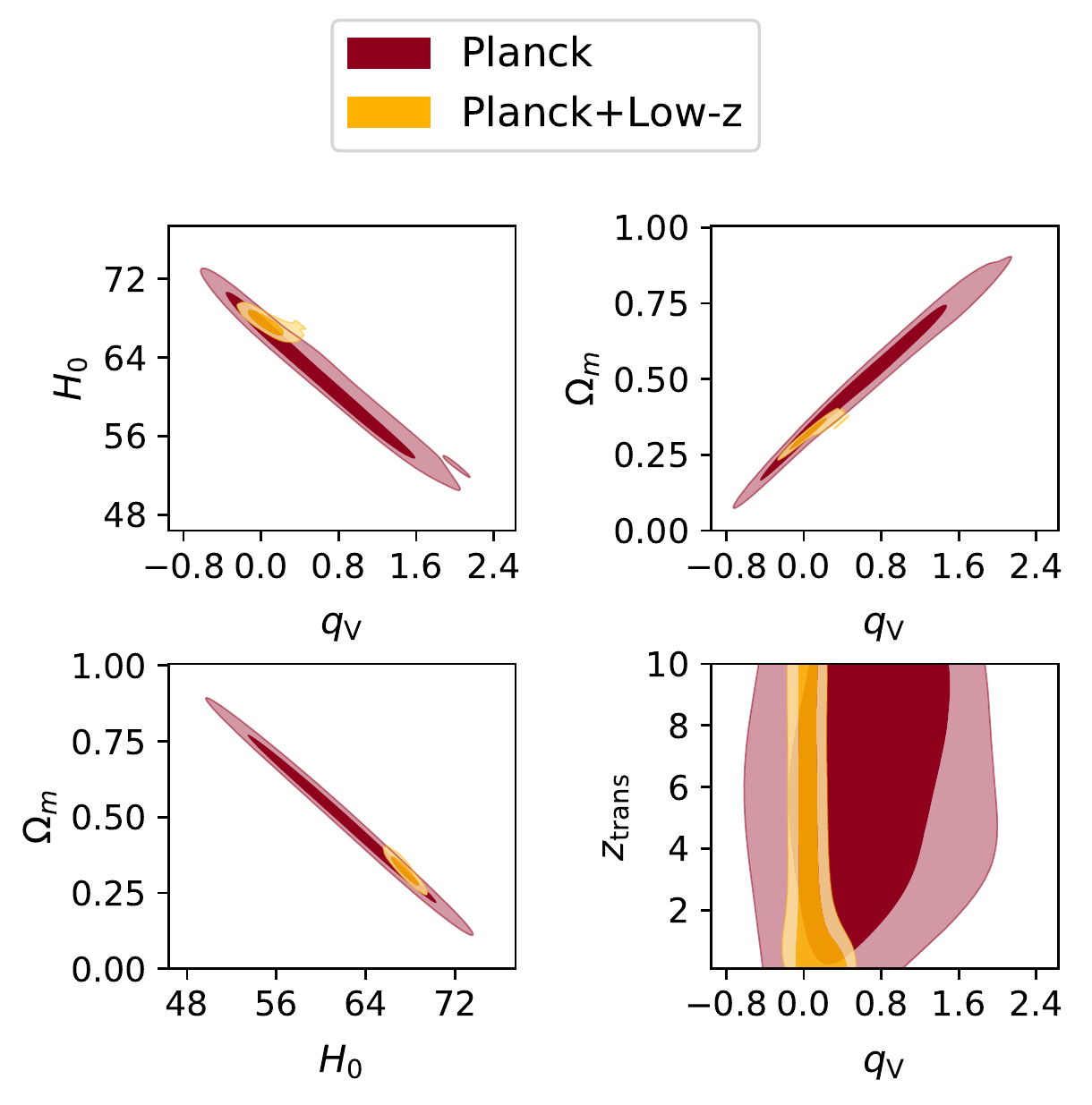}
\caption{{\it Cvar} case: $68$\% and the $95$\% confidence level marginalized contours on $H_0$, $q_{\rm V}=q_{\rm V}(z\leq z_{\rm trans})$, $z_{\rm trans}$ and $\Omega_m$ as obtained in the analysis with the Planck (red) and Planck + Low-z (yellow) datasets. \label{fig:trivar_varbin}}
\end{figure}

In Figure \ref{fig:trivar_varbin} we show the results of the case where the transition redshift $z_{\rm trans}$ is allowed to vary. In this case we also find the $\Lambda$CDM limit to be a good fit to the data, both in the Planck and Planck + Low-z combinations respectively, as reported in Table \ref{tab:constres}. We find an evolution similar to both {\it Cfix} cases, with the inclusion of the Low-z dataset breaking the degeneracies between $q_{\rm V}$ and the cosmological parameters in the Planck result. With both Planck alone and Planck+Low-z, we find that  $z_{\rm trans}$ is unconstrained, in contrast to a similar analysis in \cite{Salvatelli2014}.  For values of this parameter that correspond to the matter dominated era, this \textit{Cvar} case effectively reduces to the {\it Cfix} one, as $V(z)$ and consequently $q_{\rm V}$ become negligible. For low values of $z_{\rm trans}$ this case becomes extremely similar to $\Lambda$CDM, with $z_{\rm trans}=0$ acting as another $\Lambda$CDM limit of the model for any value the coupling can take.

\subsection{{\it SVE} case}
\begin{figure}
\includegraphics[width=.5\textwidth]{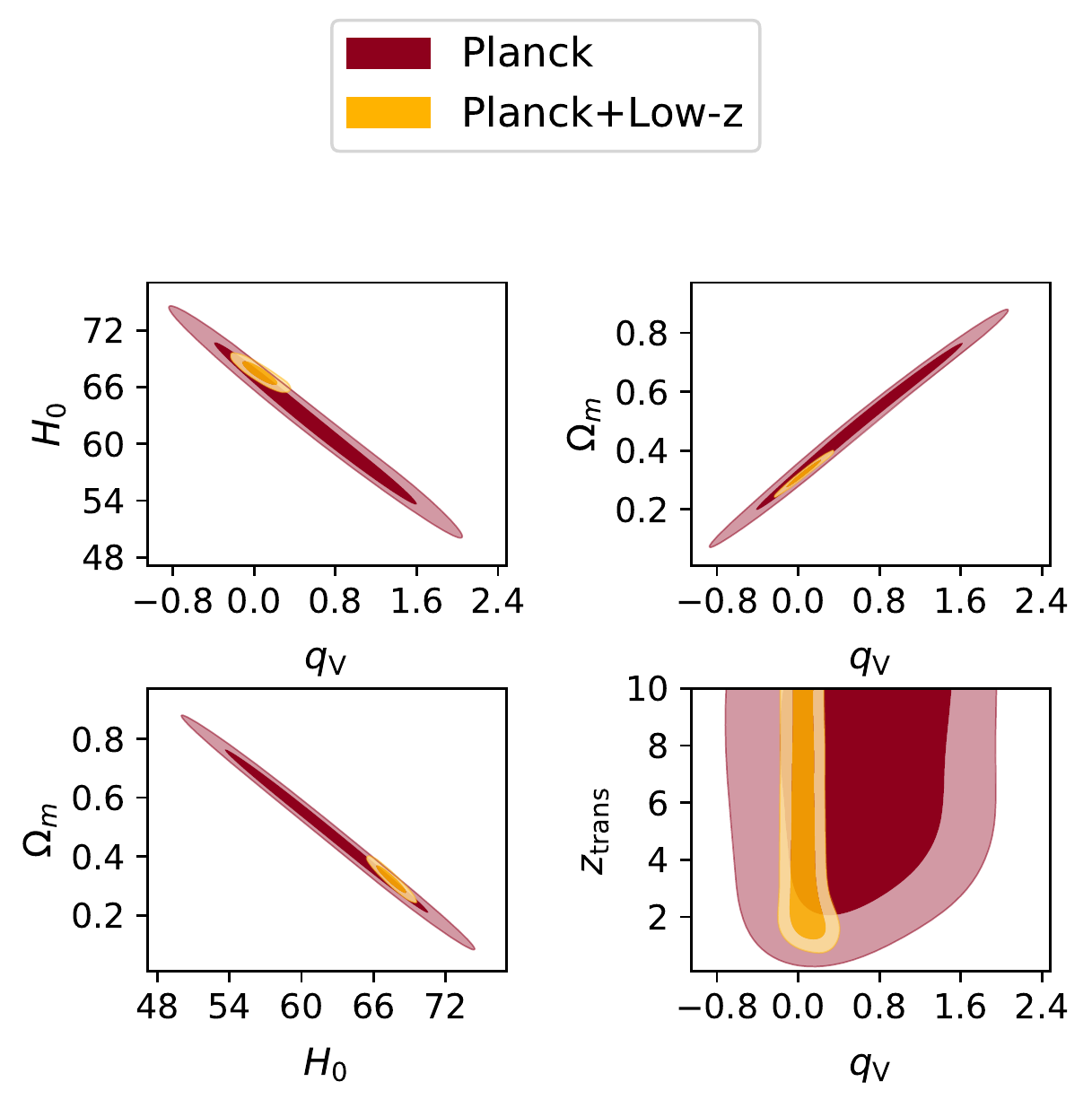}
\caption{{\it SVE} case: $68$\% and the $95$\% confidence level marginalized contours on $H_0$, $q_{\rm V}=q_{\rm V}(z\leq z_{\rm trans})$, $z_{\rm trans}$ and $\Omega_m$ as obtained in the analysis with the Planck (red) and Planck + Low-z (yellow) datasets. \label{fig:trivar_reverse}}
\end{figure}

In Figure \ref{fig:trivar_reverse}, we show the results for the {\it SVE} cosmology. The first thing to notice is that this case is analogous to {\it Cvar} when $z_{\rm trans}$ takes high values, with both data combinations favouring positive values of the coupling, i.e. a decay of vacuum energy density into CDM. This is due to the fact that in {\it Cvar}, even though $V(z)$ does not vanish, it becomes negligible in the past following the $\Lambda$CDM evolution (see Figure \ref{fig:density_ratio}) and the difference between the two models effectively vanishes. The situation is different for low transition redshifts; while in the {\it Cvar} case the model approaches $\Lambda$CDM, in \textit{SVE}, low values of this parameter are significantly disfavoured. This is because for $z_{\rm trans} \lesssim 2$, a vanishing $V(z)$ affects both the predictions for Low-z and for CMB, through its impact on CMB lensing and ISW effect. In the \textit{Cvar} case $z_{\rm trans}$ was unconstrained, while here we find a lower limit at $95\%$ confidence level of $z_{\rm trans}= 1.8$ (Planck) and $ z_{\rm trans}=1.4$ (Planck+Low-z).

\subsection{{\it 4bins} case}\label{sec:4bins}
\begin{figure}
\includegraphics[width=.5\textwidth]{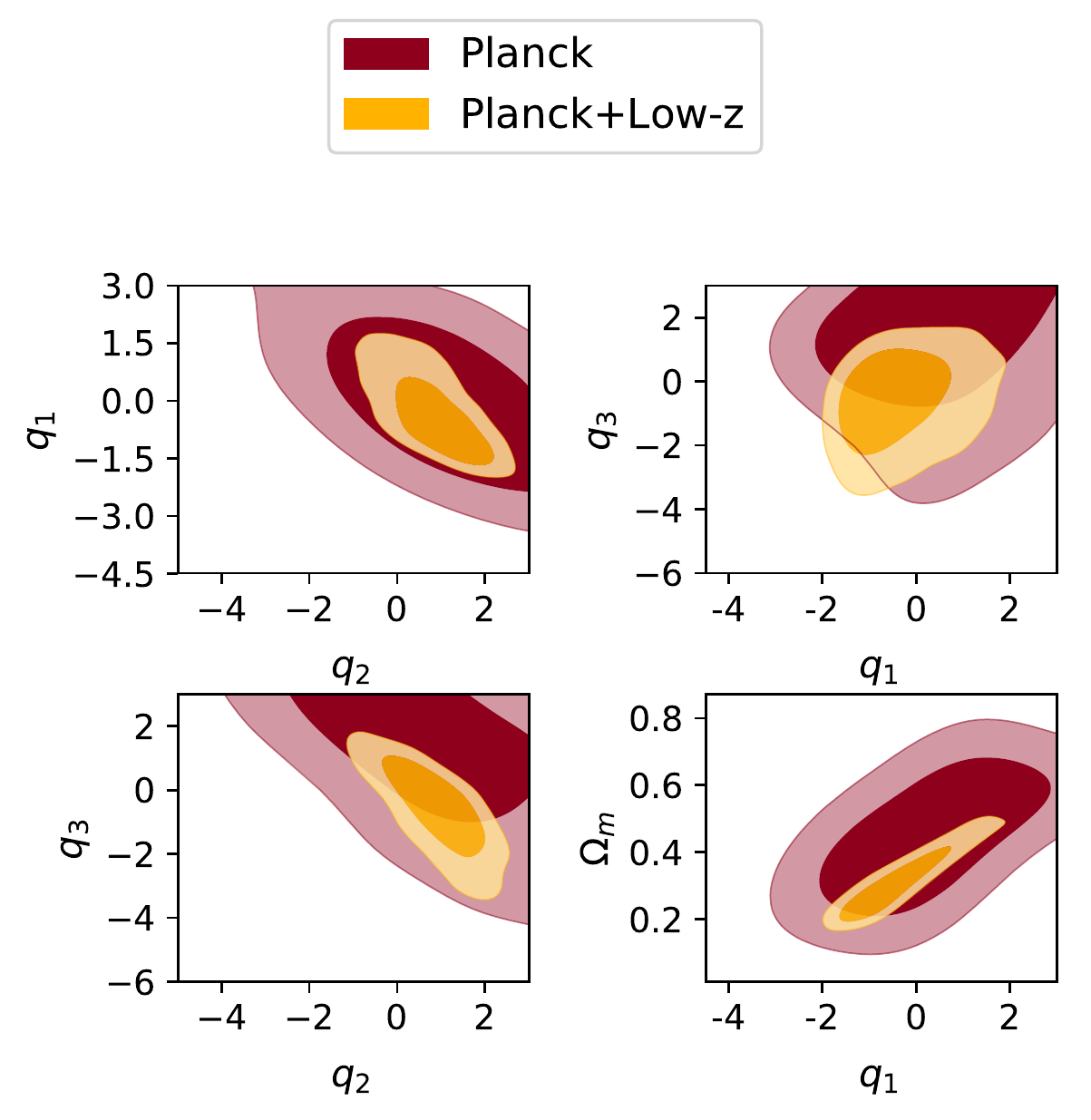}
\caption{{\it 4bins} case: $68$\% and the $95$\% confidence level marginalized contours on $q_i, \, i=1,..., 3$ and $\Omega_m$ as obtained in the analysis with the Planck (red) and Planck + Low-z (yellow) datasets.\label{fig:trivar_4bins}}
\end{figure}

In this case, we aim to update the work of \cite{Salvatelli2014}, in which the coupling consists of $N = 4$ bins in redshift, with transitions at $z = 0.3, \ 0.9, \ 2.5$ and $10$ and values $q_i$ with $i = 1, ..., 4$, thus allowing for a general evolution in redshift of the coupling function $q_{\rm V}(z)$. In Figure~\ref{fig:trivar_4bins} and Table \ref{tab:constres} we show the results obtained from the cosmological analysis with this 4 bins setup, considering both the Planck and Planck + Low-z datasets.

The first thing to note is that the high redshift bin $q_4$ is not constrained by either dataset. This is due to the fact that most of the Low-z data lie at redshifts lower than those affected by this parameter and therefore any constraining power would come from the effect of the coupling in this redshift bin on CMB power spectra predictions. However, we see that the Planck data is also unable to place any bounds on the value of $q_4$, nor an upper bound on the value of $q_3$.

While $\Lambda$CDM is also a good fit to the data in this case, in general we find that the allowed range for the amplitude of the interaction in each redshift bin is larger than in the {\it Cfix} and {\it Cvar} cases. This is expected, as the values of $q_i$ can be compensated for by the overall evolution of $q_{\rm V}(z)$ and therefore by the $q_{j\neq i}$ parameters. This induces an anti-correlation between the values of the coupling in neighbouring bins. Once again, this degeneracy is significantly reduced when the Low-z data are included, as these datasets are more efficient in constraining the values of $q_i$ in each redshift bin rather than the average effect of the interaction.

However, while in the {\it Cfix} and {\it Cvar} cases the inclusion of Low-z produces tight posteriors centered on the $\Lambda$CDM limit, in the \textit{4bins} case the first bin posterior is slightly shifted to negative values (with $q_1=0$ still within the 68\% confidence interval) and the second bin posterior is shifted towards positive values: this is due to the aforementioned anti-correlation. While still in agreement with a constant $q_{\rm V}(z)=0$ cosmology, the Planck+Low-z dataset allows for a model with an oscillatory amplitude of vacuum energy-CDM interaction at low redshifts (See Section \ref{subsec:gpres} for further discussion). This is in contrast to the results of many similar works. We will expand on this point in Section \ref{sec:rebuttal}.

\subsection{Evolution of $f \sigma_8$}
\begin{figure}
\includegraphics[width=.5\textwidth]{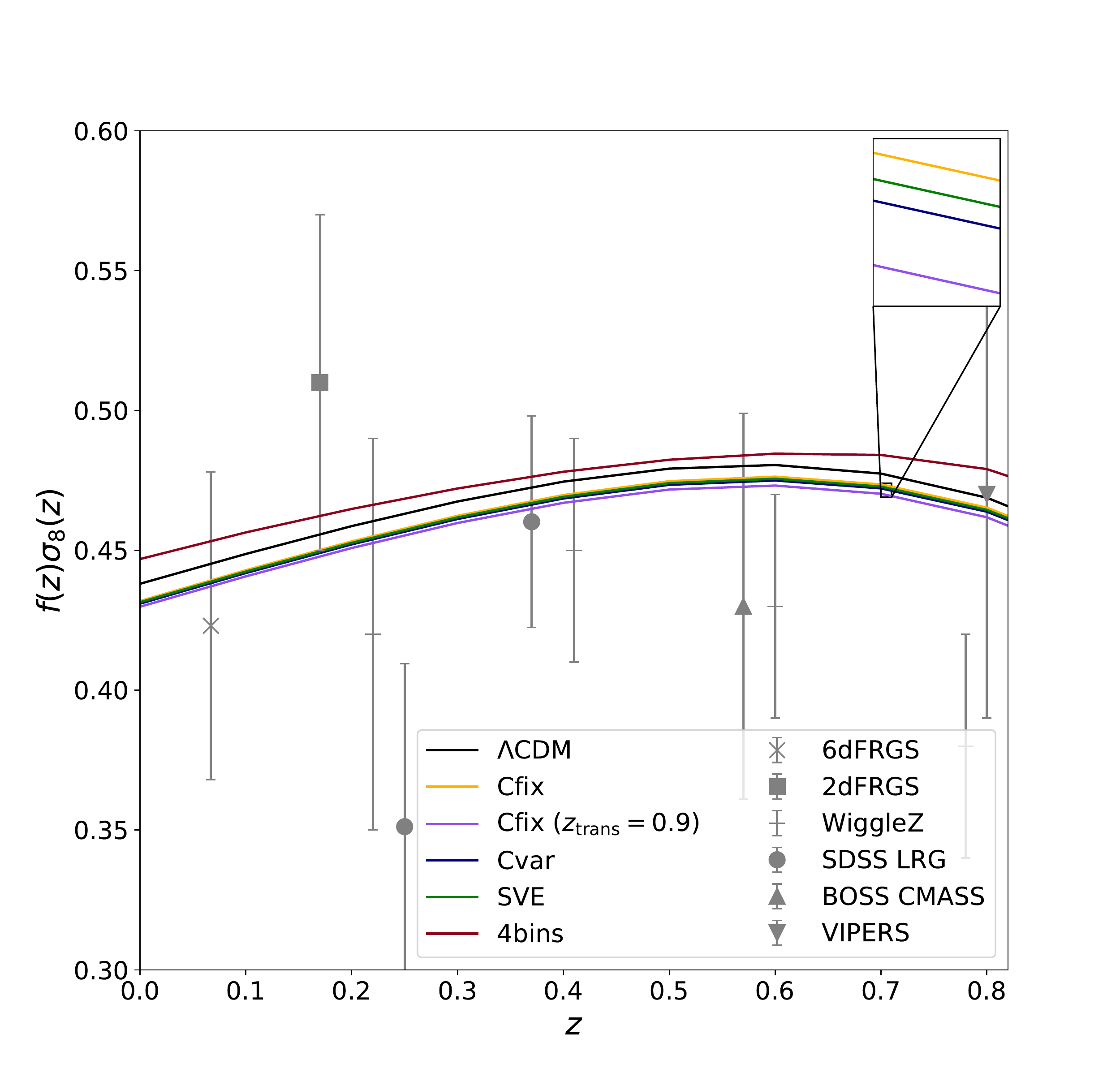}
\caption{The predictions for $f \sigma_8$ for \LCDM~(plotted in black) and the interacting cosmologies studied in this work. For illustrative purposes, we plot these together with data from various collaborations (see text for details).\label{fig:fsigma8}}
\end{figure}

From these results, we can also examine how the interaction in each case affects the evolution of the $f \sigma_8$ parameter as computed by the modified {\tt CAMB}, keeping in mind that in our interacting scenario this parameter does not directly constrain the growth factor, i.e.\ it rather represents $f_i \sigma_8$, as discussed in subsection \ref{subsec:rsd}. In Figure \ref{fig:fsigma8}, we plot the $f\sigma_8$ prediction for each case, using the mean posterior values of $q_{\rm V}$ from the Planck+Low-z runs to obtain its evolution as a function of redshift. For illustrative purposes, we plot these predictions along with data points from various collaborations: 2dFGRS \protect\citep{2df_fs8}, 6dFGRS \protect\citep{6df_fs8}, WiggleZ \protect\citep{wigglez_fs8}, SDSS LRG \protect\citep{SDSS_fs8}, BOSS CMASS \protect\citep{boss_fs8} and VIPERS \protect\citep{vipers_fs8}. 

This plot shows how the similar values of $q_{\rm V}$ obtained for \textit{Cfix}, \textit{Cvar} and \textit{SVE} lead to similar evolution histories for $f\sigma_8$, with the small positive values of $q_{\rm V}$ in these cases leading to a suppression of this quantity with respect to $\Lambda$CDM. Growth is suppressed with a positive coupling because our implementation in \texttt{CAMB} works by starting with the values of cosmological parameters at $z=0$ and evolving them backwards in time. This means that, with a positive $q_{\rm V}$, we need less matter in the past to reach the correct value of $\Omega_m$ today; in addition, $q_{\rm V}>0$ implies a negative contribution of the coupling  to $\dot{\delta}$ in \eqref{eq:densitycontrast}; the net result is that the growth is suppressed. The \textit{4bin} case instead sees an enhancement of $f\sigma_8$ with respect to $\Lambda$CDM: this is due to the overall negative value of the coupling across the 4 redshift bins.

Note that for $q_{\rm V}\not = 0$, Figure \ref{fig:fsigma8} is effectively a plot of $f_i \sigma_8$, and $f_i>f$ for $q_{\rm V}>0$ (see Eq.\ \eqref{eq:fi}). In practice, the suppression of the growth implies a $\sigma_8$ small enough to produce a smaller $f_i \sigma_8$, and vice versa for  $q_{\rm V}<0$.

\section{Discussion} \label{sec:discussion}
In this section we discuss our results, presenting a rough model comparison analysis in order to estimate the statistical preference of our models with respect to $\Lambda$CDM. Moreover, we focus on the effects on the tensions in the values of $H_0$ and $\sigma_8$ in the different interacting cases presented above. We also describe how the $q_{\rm V}(z)$ function can be reconstructed using Gaussian processes.

\subsection{Model comparison}
In all our results we find a good agreement between the $\Lambda$CDM limit of the interacting models investigated and the constraints obtained through the analysis of cosmological data. We therefore expect that there is no significant statistical preference for the extended model over $\Lambda$CDM. However, we will quantify this preference by making use of the {\it Deviance Information Criterion} (DIC) \citep{RSSB:RSSB12062}:
\begin{equation}
\text{DIC}\equiv\chi^2_{\rm eff}(\hat{\theta})+2p_D \,,
\end{equation}
where $\chi^2_{\rm eff}(\hat{\theta})= -2\ln{\mathcal{L}(\hat{\theta})}$, $\hat{\theta}$ is the parameter vector at the best fit and $p_D=\overline{\chi^2_{\rm eff}(\theta)}-\chi^2_{\rm eff}(\hat{\theta})$, where the bar denotes the average taken over the posterior distribution.
This estimator accounts for both the goodness of fit through $\chi^2_{\rm eff}(\hat{\theta})$ and for the Bayesian complexity of the model, $p_D$, which disfavours models with extra parameters.
In order to compare $\Lambda$CDM with the models explored here, we compute:
\begin{equation}
\Delta\text{DIC} = \text{DIC}_{\text{V}} - \text{DIC}_{\Lambda \text{CDM}}.
\end{equation}  
From this definition it follows that a negative $\Delta\text{DIC}$ would support the extended model, while a positive one would support $\Lambda$CDM.

In Table \ref{tab:DIC} we show the values obtained for this estimator in all the cases analyzed in this paper. We find that when analyzing only CMB data, all the models except for {\it Cfix} are slightly preferred with respect to $\Lambda$CDM. However, all the cases have a $\Delta$ DIC close to zero, showing that the preference of the extension over the standard model (or vice versa) is inconclusive in all cases, if we set $\Delta \text{DIC} = 5$ as the threshold for a moderate preference \citep{Joudaki:2016kym}. When analyzing the Planck+Low-z case, we find that all cases have a small positive $\Delta$ DIC, indicating that $\Lambda$CDM is marginally preferred over the extended model. This comes from the fact that adding the Low-z datasets significantly shrinks the constraints around the $\Lambda$CDM limit of the model, thus disfavouring the extended case which, at this point, effectively reproduces a $\Lambda$CDM cosmology with the addition of extra parameters.

\begin{table}
\centering
\begin{tabular}{|c|c|c|}
\hline
Parameter                             & Planck   &  Planck+Low-z \\
\hline
\textit{Cfix}                         &  $1.1$   & $3.8$\\
\textit{Cfix ($z_{\rm trans} = 0.9$)} &  $-1.2$  & $0.4$\\
\textit{Cvar}                         &  $-0.5$  & $2.6$\\
\textit{SVE}                          &  $-1.3$  & $1.3$\\
\textit{4bins}                        &  $-1.6$  & $3.1$\\
\hline
\hline
\end{tabular}
\caption{$\Delta DIC$ values for the different models analyzed, both when using Planck data alone and when combining them with the Low-z datasets.}\label{tab:DIC}
\end{table}

\subsection{Effects on cosmological tensions}
\begin{figure*}
\includegraphics[width=.4\textwidth]{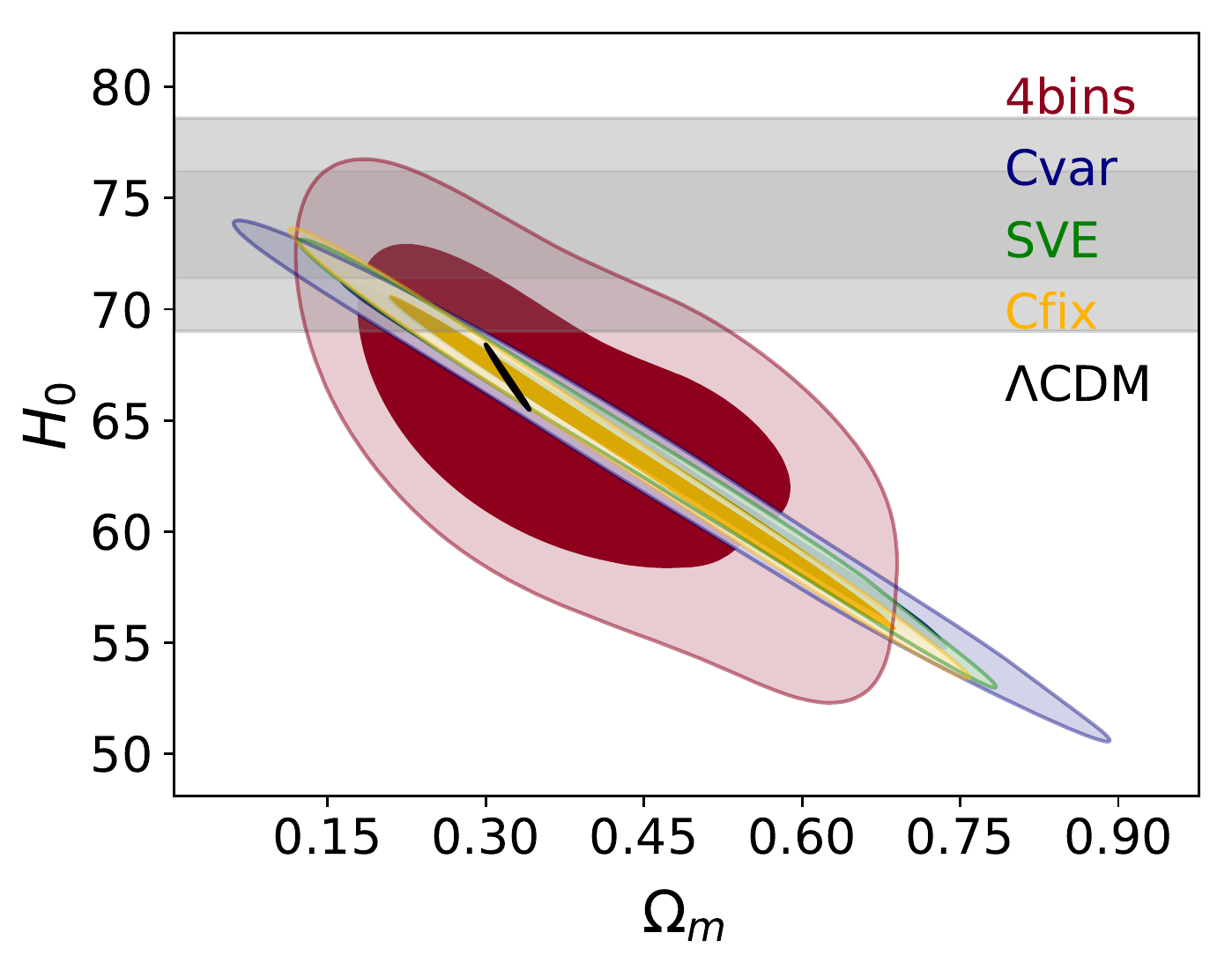}
\includegraphics[width=0.4\textwidth]{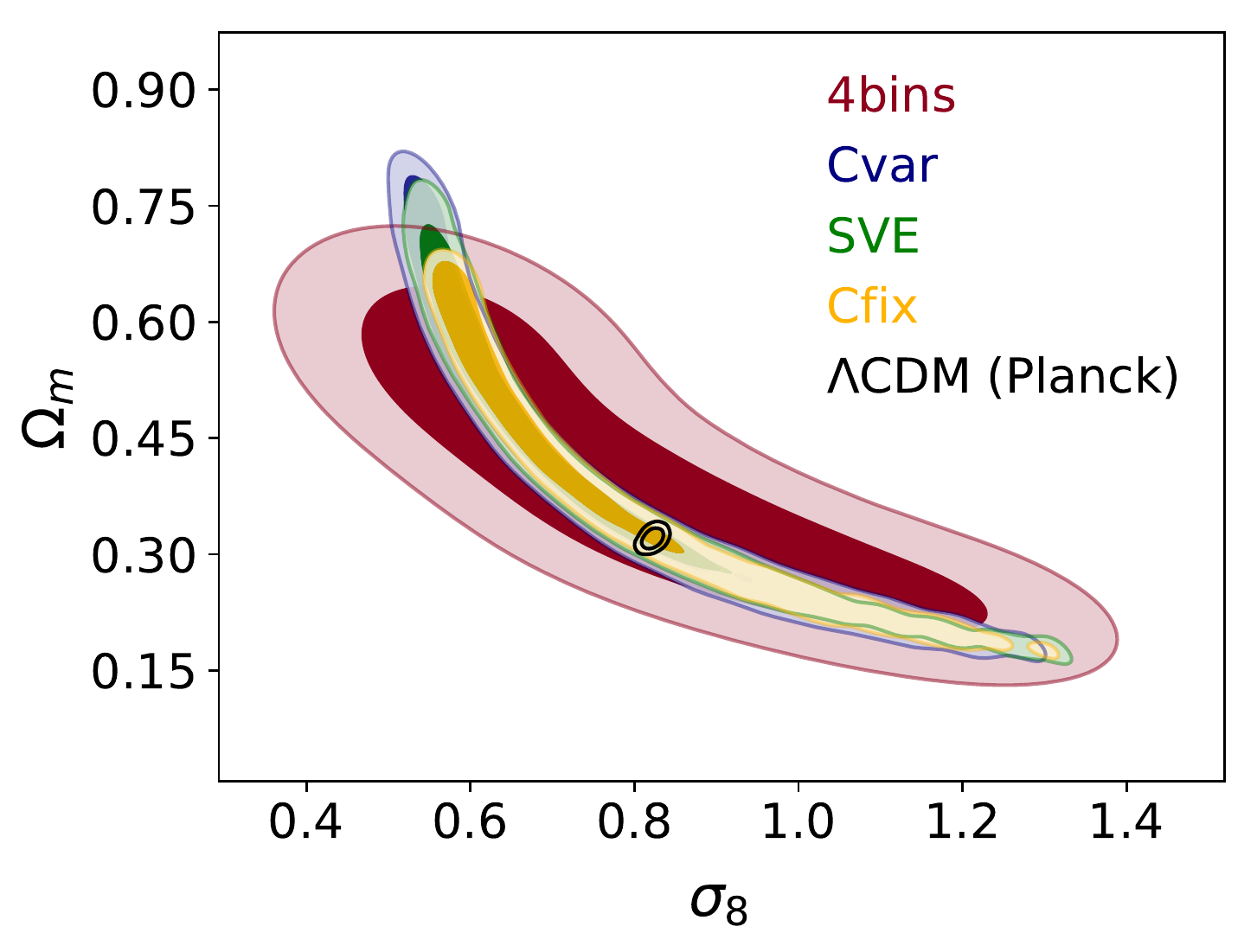}\\
\caption{$68\%$ and $95\%$ confidence levels on the $H_0$ -- $\Omega_m$ plane (left panel) and $\Omega_m$ -- $\sigma_8$ plane (right panel) for the 4 cosmologies considered: {\it Cfix} (yellow contours), {\it Cvar} (dark blue contours), {\it 4bins} (red contours) and {\it SVE} (green contours), with the $\Lambda$CDM Planck alone case plotted in black. The grey bands in the left panel show the $68\%$ and $95\%$ confidence level on $H_0$ as obtained in \protect\cite{Riess2018}. These results are obtained with the analysis of the full Planck dataset.\label{fig:tension1}}
\end{figure*}

As we highlighted in Section \ref{sec:intro}, one of the motivations to explore the coupling scenarios discussed in this paper is to attempt to solve the tensions that exist between different observations, i.e.\ the discrepancies between low and high redshift measurements of the present day expansion rate of the Universe and of the clustering of matter. In Figure \ref{fig:tension1} we plot the $H_0$ vs $\Omega_m$ and $\sigma_8$ vs $\Omega_m$ 2D marginalized contours for every case considered, obtained using the Planck 2015 dataset, comparing them with the constraints used assuming $\Lambda$CDM, in order to examine the effects of the interaction on the $H_0$ and $\sigma_8$ tensions.

\begin{figure*}
\includegraphics[width=\textwidth]{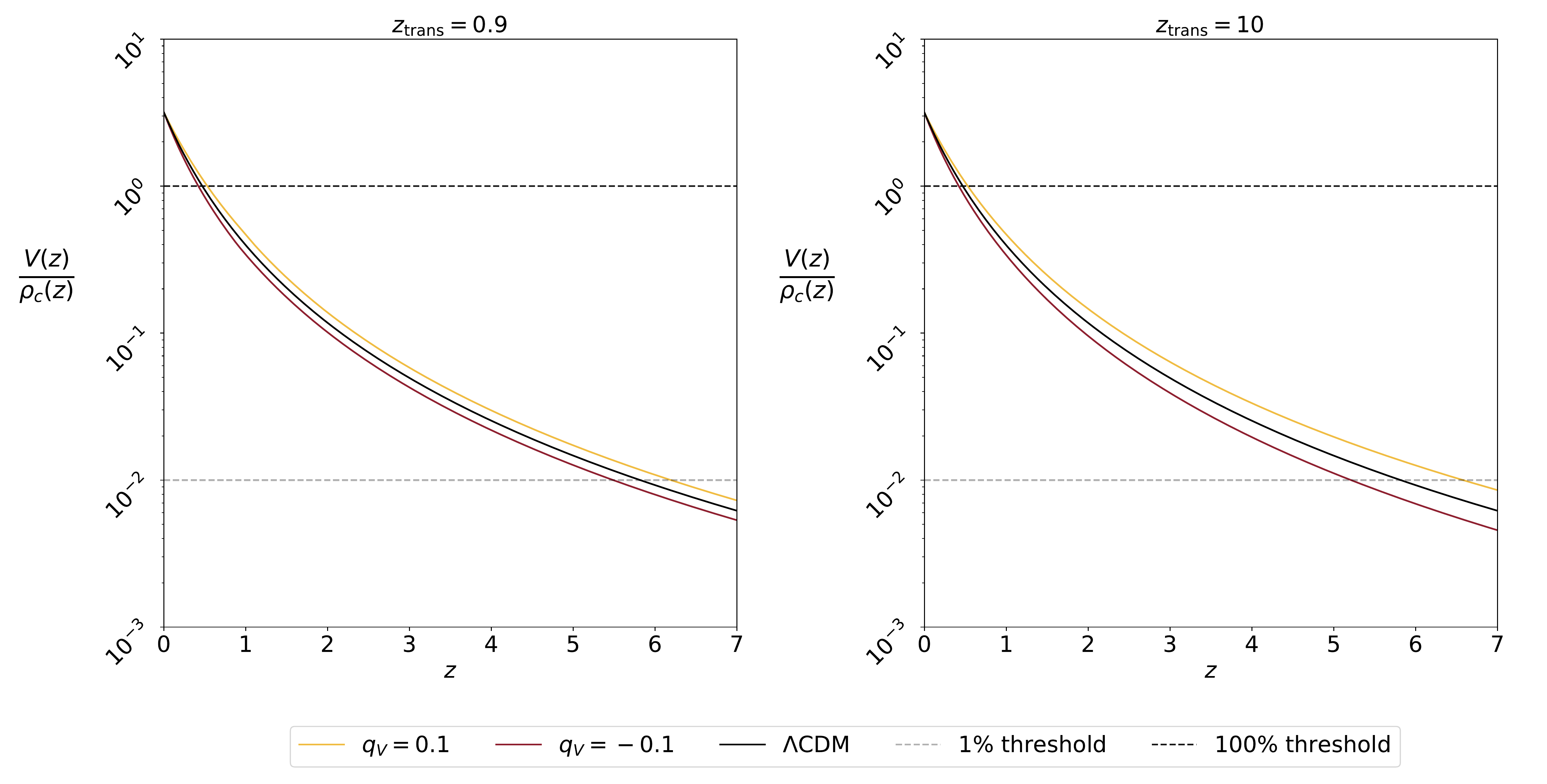}
\caption{Ratio of the vacuum to CDM energy density for a small positive and negative coupling with two different transition redshifts. The \LCDM~case is plotted in dark blue.\label{fig:density_ratio}}
\end{figure*}

We firstly note that for both of these combinations, the contours obtained for the {\it Cfix}, {\it Cvar} and {\it SVE} are very similar, showing that changing the behaviour of $V(z)$ after $z_{\rm trans}$ (from standard $\Lambda$CDM evolution to vanishing $V(z)$) has no significant effect if $z_{\rm trans}$ is already in an epoch where vacuum energy is negligible. In Figure \ref{fig:density_ratio}, we have plotted the ratio of the vacuum to CDM energy densities, for both a small positive and negative coupling and with two transition redshifts, $z_{\rm trans} = 0.9 \text{ and } 10$. The sign of the coupling and the transition redshift value have limited effect, as for each of the four values shown, the density ratio reaches $1/100$ and $1$ at very similar redshifts. The {\it 4bins} case instead yields broader constraints with respect to the other cases, an effect which is due to the higher number of coupling parameters and their degeneracies with the standard cosmological ones.

The left panel of Figure \ref{fig:tension1} shows how the coupling scenarios are able to apparently ease the tension between the local measurements of $H_0$ (grey band) and the Planck measurement. However, this is only due to the extreme degeneracy between $H_0$, $\Omega_m$ and $q_{\rm V}$ that we highlighted in Section \ref{sec:results}; the mean values obtained for $H_0$ are actually lower than those found by Planck assuming $\Lambda$CDM, and the tension is eased only because of the much larger error bars. In \cite{Poulin:2018cxd} it was proposed that this tension could be relaxed with an Early Dark Energy component, affecting the evolution of the Universe at $z\gtrsim3000$; while not explored here, a high redshift coupling between CDM and vacuum energy could in principle be used to mimic the effect of such a component. We leave the investigation of this possibility for a future work.

In the right-hand panel of Figure \ref{fig:tension1}, we instead highlight how reconciling the tension in $\sigma_8$ is less feasible in this model. The errors on the cosmological parameters are once again enlarged by the degeneracies introduced by the coupling. This leads to lower values of $\sigma_8$ being allowed, but these lower values subsequently necessitate higher values of $\Omega_m$ in compensation, which are then disfavoured by the Low-z data.

\subsection{Gaussian process reconstruction}\label{subsec:gpres}
We can use Gaussian processes to attempt to reconstruct the \qv~function for the 4 bin case. Gaussian processes have been widely used in cosmology to reconstruct smooth functions from observational data, particularly for functions such as $H(z)$ and the dark energy equation of state $w(z)$ (see, for example, \cite{Seikel2012,Shafieloo2012,Cai:2015zoa,Zhang2018}). Since we do not expect the \qv~function to vary rapidly, the GP approach is suitable to use in this case too. We use the Gaussian process regressor available in the Python library \texttt{george}\footnote{\url{https://github.com/dfm/george}}. 

The Gaussian process regression works by using a covariance function, or kernel, to relate the function values at two points, $x$ and $\tilde{x}$, to each other. The advantage of using Gaussian processes over a basic spline or parametric fit is that it not only allows us to consider a much wider range of possible fitting functions for \qv~but it also means we can potentially inform our choice of kernel based on the underlying physical processes at work. 

There has been some debate in the literature about the appropriate choice of kernel for various problems, with no clear-cut answer yet. For example, \cite{Seikel2013} found that the Mat\'{e}rn class of kernels, and especially the Mat\'{e}rn ($\nu = 9/2$) kernel was the most successful at reconstructing $w(z)$ using supernova data. The Mat\'{e}rn class of kernels have the following general form
\begin{align}
k(x, \tilde{x}) = \sigma^2 \frac{2^{1- \nu}}{\Gamma(\nu)} \left(\frac{\sqrt{2 \nu (x - \tilde{x})^2}}{\ell} \right)^\nu \times K_\nu \left(\frac{\sqrt{2 \nu (x - \tilde{x})^2}}{\ell} \right), \label{eq:matern}
\end{align}
where $\Gamma(\nu)$ is the gamma function, $K_\nu$ is a modified Bessel function and $\nu$ controls the shape of the covariance function, tending to the Gaussian limit as $v\rightarrow \infty$. The hyperparameters $\ell$ and $\sigma$ correspond to the approximate length scale over which the function varies and the magnitude of these variations respectively.

\begin{figure}
\includegraphics[width=0.5\textwidth]{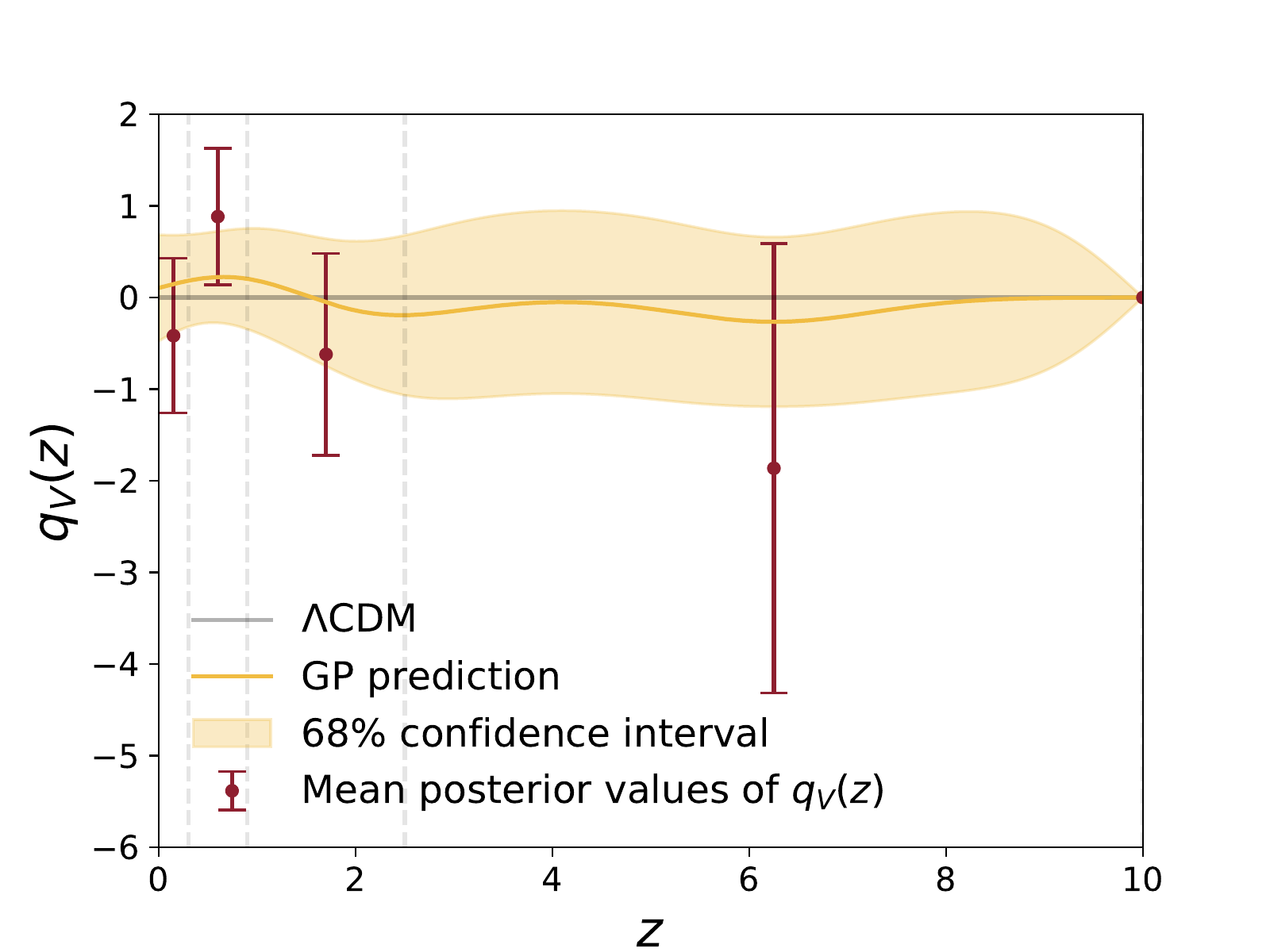}
\caption{Gaussian process reconstruction of \qv~using the squared exponential kernel, with data points as provided by the analysis of the {\it 4bins} cosmology. The grey dashed lines indicate the boundaries of each redshift bin.\label{fig:4binGP}}
\end{figure}

\begin{figure}
\includegraphics[width=0.5\textwidth]{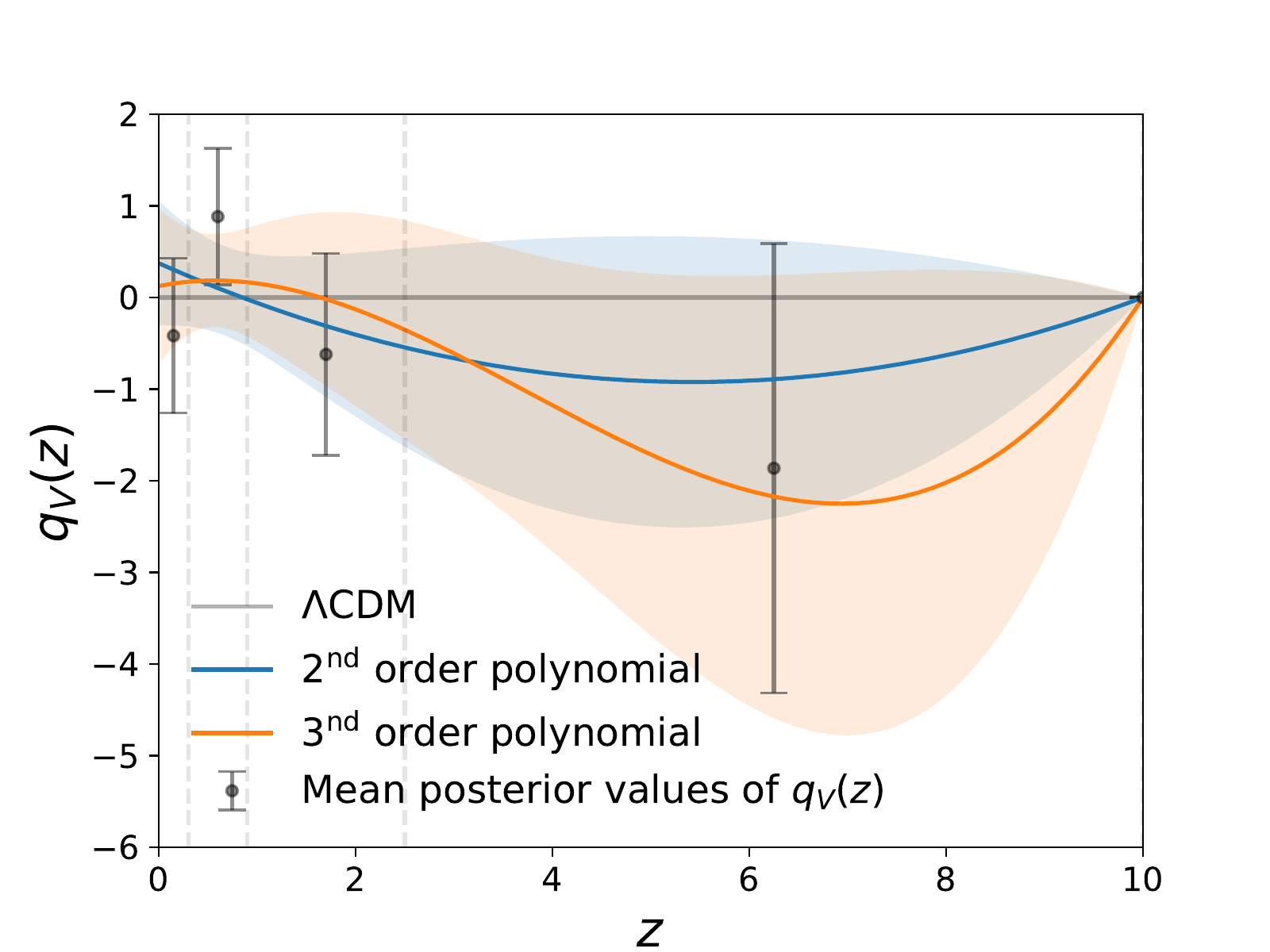}
\caption{Gaussian process reconstruction of \qv~using three different kernels, with data points as provided by the analysis of the {\it 4bins} cosmology. The grey dashed lines indicate the boundaries of each redshift bin and the shaded regions denote the 68\% confidence intervals of the GP reconstruction. \label{fig:GPcomparison}}
\end{figure}

In the course of our analysis we investigated the results given by all the basic kernels provided by \texttt{george}, none of which resulted in a function that excludes \LCDM~ at any confidence level, but as kernels can be added or multiplied in almost any combination, we did not test every possibility exhaustively. We therefore present the reconstruction given by the squared exponential kernel, the simplest of the Mat\'{e}rn class kernels, recovered from \eqref{eq:matern} when $\nu \rightarrow \infty$,
\begin{align}
k(x, \tilde{x}) = \sigma^2 \exp\left(-\frac{(x - \tilde{x})}{2 \ell^2}\right).
\end{align}
This reconstruction is shown in Figure \ref{fig:4binGP}. For comparison, we also show the reconstruction using the \nth{2} and \nth{3} order polynomial kernels provided by \texttt{george} in Figure \ref{fig:GPcomparison}. The data points in both cases come from the Planck + Low-z runs, in which we can clearly see the oscillatory behaviour of the coupling mentioned earlier.

The hyperparameters $\ell$ and $\sigma$ that appear in the kernels described above can be optimized by maximizing the log-likelihood of the functions they produce. However, with this optimization implemented, our GP regressions all collapsed to be exactly equal to zero for all redshifts. This is because we have very little data with which to inform the Gaussian process and the GP always returns to its baseline of zero when it has insufficient information. We therefore conclude that the GP will be better suited to reconstructing a case with many more redshift bins, which we intend to investigate in a future work.

\section{Comment on other results} \label{sec:rebuttal}
Finally, we note that there has been an extensive treatment in the literature of a similar interacting vacuum scenario to that studied in this work \citep{Kumar:2017dnp, Sola:2017znb, Sola:2017jbl, Tsiapi:2018she, Kumar:2019wfs}, upon which we would like to comment.

Firstly, all of the aforementioned works appear to use a single bin case, akin to what we call \textit{Cfix}, which implies the interaction parameter $q_{\rm V}$ has been constant throughout the entire cosmic history. This is sufficient for a basic analysis, but carries some important physical implications. If the interaction remains constant for the entire cosmic history (and is found to favour a decay of CDM into the vacuum) it implies that eventually the energy density of CDM must become negative, as we have pointed out at the end of Section \ref{subsec:const}. While the phenomenology of such a scenario may still be interesting when studying the Universe's history, the unphysicality of the model is motivation enough to instead consider the effects of a dynamical interaction, as we have done in this work.

Secondly, in \cite{Tsiapi:2018she}, the effect of the interaction on perturbations in the matter energy density are not clearly taken in to account. As we have shown, the interaction enters into the equation for the density contrast \eqref{eq:densitycontrast} and it is necessary to modify \texttt{CAMB} accordingly. The presence of the coupling in this equation means that the interaction will have some effect on cosmological structure growth, which is also clear from the matter power spectrum for the \textit{Cfix} case, as shown in Figure \ref{fig:matterpower}. While an analysis of the background cosmology is instructive, we consider our current work an improvement, as we also take into account the effect of the coupling at the level of the perturbations. 

Furthermore, \cite{Sola:2017znb} and \cite{Sola:2017jbl} forgo a complete MCMC parameter inference with the full Planck CMB likelihood, choosing instead to use only the compressed likelihood. We believe our full analysis that takes the complete, uncompressed data into account has produced a more reliable result. However, we note that the novel use of the bispectrum as a potential tracer of the dynamics of dark energy was investigated in \cite{Sola:2017jbl} and subsequently expanded on in \cite{Sola2018:lcd} using the well-known XCDM, CPL and $\phi$CDM parametrizations (in this work the authors also used the full Planck likelihood). Such an idea was also proposed by \cite{Borges:2017jvi}, but we emphasise that the original arXiv version of \cite{Sola:2017jbl} preceded \cite{Borges:2017jvi}.  It was indeed found in \cite{Sola2018:lcd} that the bispectrum enhances the dynamical dark energy signal, so an interesting avenue of future investigation would be to use the bispectrum data when constraining the interacting vacuum scenario. A careful consideration of the effect of the interaction on the bispectrum measurements would be needed, however.

The work of \cite{Kumar:2017dnp} also examined the same interacting scenario, but in addition to varying the interaction strength along with the six standard cosmological parameters in \LCDM, they also varied the sum of the neutrino masses, $\sum m_\nu$, and the effective number of relativistic degrees of freedom, $N_{\rm eff}$. It was found in that work that the use of the combination of Planck+BAO+JLA data (exactly equivalent to the Planck+Low-z combination used in this work) resulted in finding no suggestion of an interaction. However, the inclusion of galaxy cluster count data from Planck \citep{PlanckSZ2015} and CFHTLenS \citep{Heymans:2013fya} resulted in finding a non-zero interaction at the 99\% confidence level. 

In \cite{Kumar:2019wfs}, the authors again tested the same interacting scenario, but with yet another combination of datasets: Planck 2015 with the KiDS weak lensing survey \citep{Kohlinger2017} and the 2016 Hubble Space Telescope measurement of $H_0$ \citep{Riess2016}. In this work, the authors found compelling statistical evidence for an interaction and were also able to simultaneously relax the $H_0$ and $\sigma_8$ tensions. This again indicates the strong effects that different datasets can have and demonstrates the need for awareness of possible systematics when choosing and combining datasets. In particular, when using weak lensing data, it important to make a conservative cut of the non-linear scales in these datasets, unless the non-linear theory for perturbations is known.

Finally, we would like to address some differences between this work and the previous work of some of the current authors \citep{Salvatelli2014}. The current work was partly designed to make a comparison with the work presented in that Letter, confronting the same interacting scenario with the latest available datasets. In \cite{Salvatelli2014}, it was found that a late-time interaction in a single low redshift bin of $z\leq 0.9$ was favoured over the null interaction case, with \LCDM~being excluded at 99\% confidence level. As described in subsection \ref{subsec:Low-z_cfix}, we replicated this  case, \textit{Cfix} with $z_{\rm trans}=0.9$, albeit using more up-to-date datasets (the Planck 2015 likelihood and newer BAO, RSD and Type Ia supernovae data), as well as a broader prior on the parameter $q_{\rm V}$ that includes positive values,  but found no significant deviation from \LCDM~at all. 

Similarly, when replicating the 4 bin case, also analysed by \cite{Salvatelli2014}, we found no significant deviation from \LCDM~at low redshift, in contrast to the 95\% confidence level difference reported in that work. We can possibly attribute this to the simple lack of evidence for an interaction in the newer observational datasets used in the current work. Our finding that the null interaction scenario (i.e.\ \LCDM) is always well within the 95\% confidence region for $q_{\rm V}$ is in agreement with the recent work by \cite{Yang:2018qec}.

\section{Conclusions} \label{sec:conclusions}
In this work we have considered the possibility of an interaction in the dark sector, represented as a pure energy exchange between vacuum and cold dark matter. We have investigated constraints on this scenario, by making a simple binned parametrization of the coupling function in redshift, using the latest cosmological datasets to place constraints on the coupling in each bin. 

We investigated a number of different cases under the umbrella scenario of the interacting vacuum, namely the cases with a single bin and either a fixed or varying transition redshift (\textit{Cfix} and \textit{Cvar}); a case in which the vacuum energy is zero at early times, only growing after the interaction switches on, and lastly, in a model-independent way, a four bin case to replicate the work of \cite{Salvatelli2014}.

In all the cases we studied, we found that the \LCDM~case, corresponding to no interaction in our scenario, is always well within the 95\% confidence regions of our parameter estimation. At the same time the interacting scenario remains a viable alternative to $\Lambda$CDM, and only future data will be able to settle the case. We also note that our analysis is restricted to linear scales, while it is entirely possible that in extending the interacting vacuum scenario to non-linear scales more stringent constraints will be found, cf.\ \cite{He:2018oai}. Our findings are in contrast to a number of recent works mentioned in the previous section, but we have described the differences in our approach and contest that these  are sufficient to explain the different results.

Finally, we note that the observational literature is being continually updated, with ever-larger surveys and telescopes planned for the near future. With these surveys will come an unprecedented level of precision in the measurement of cosmological observables that will in turn demand the utmost rigor from models designed to predict their values. The careful consideration of every implication a model may carry is therefore of paramount importance, and models that take into account the background cosmology only will no longer be satisfactory explanations of our observational data. 

\section*{Acknowledgements}
We thank Robert Crittenden and Antony Lewis for helpful comments, and Najla Said and Valentina Salvatelli for useful discussions in the early stages of this work. This paper is based upon work from COST action CA15117 (CANTATA), supported by COST (European Cooperation in Science and Technology). Numerical computations were done on both the Sciama High Performance Compute (HPC) cluster which is supported by the ICG, SEPNet and the University of Portsmouth, and the Maris cluster, which is supported by the Lorentz Institute, Leiden University. MM acknowledges support from the D-ITP consortium, a program of the NWO that is funded by the OCW. NBH is supported by UK STFC studentship ST/N504245/1 and gratefully acknowledges the hospitality of the Lorentz Institute where part of this work was carried out. SP acknowledges support from the NWO and the Dutch Ministry of Education, Culture and Science (OCW), and also from the D-ITP consortium, a program of the NWO that is funded by the OCW.  
MB and DW are supported by UK STFC Grant No. ST/N000668/1.

%%%%%%%%%%%%%%%%%%%%%%%%%%%%%%%%%%%%%%%%%%%%%%%%%%



\bibliographystyle{mnras}
\bibliography{void} % if your bibtex file is called example.bib
%\nocite{*}

%%%%%%%%%%%%%%%%%%%%%%%%%%%%%%%%%%%%%%%%%%%%%%%%%%

% Don't change these lines
\bsp	% typesetting comment
\label{lastpage}
\end{document}